\documentclass[12pt]{article}

\usepackage{amsmath,amssymb,graphicx,multirow,xspace}
\usepackage[colorlinks=true,urlcolor=blue,anchorcolor=blue,citecolor=blue,filecolor=blue,linkcolor=blue,menucolor=blue,pagecolor=blue]{hyperref}
\usepackage[compress,numbers]{natbib}
\usepackage{subfigure, placeins}
\usepackage{booktabs}
\usepackage{blindtext, rotating}
\usepackage{afterpage}
\usepackage{enumitem}
\usepackage{ marvosym }
\usepackage{verbatim}
\usepackage{authblk}

\newcommand{\newc}{\newcommand}
\newc{\gsim}{\lower.7ex\hbox{$\;\stackrel{\textstyle>}{\sim}\;$}}
\newc{\lsim}{\lower.7ex\hbox{$\;\stackrel{\textstyle<}{\sim}\;$}}
\newc{\gev}{\,{\rm GeV}}
\newc{\mev}{\,{\rm MeV}}
\newc{\ev}{\,{\rm eV}}
\newc{\kev}{\,{\rm keV}}
\newc{\tev}{\,{\rm TeV}}

\def\ln{\mathop{\rm ln}}

\def\Tr{\mathop{\rm Tr}}

\def\Re{\mathop{\rm Re}}

\newc{\mz}{M_Z}
\newc{\mpl}{M_*}
\newc{\mw}{m_{\rm weak}}
\newc{\nr}[1]{N^c_R{}_{#1}}
\usepackage{amsmath}

\def\bitem{\begin{itemize}}
\def\eitem{\end{itemize}}
\newc{\ie}{{\it i.e.}}          \newc{\etal}{{\it et al.}}
\newc{\eg}{{\it e.g.}}          \newc{\etc}{{\it etc.}}
\newc{\cf}{{\it c.f.}}

\newcommand{\CO}{\mathcal{O}}

\newcommand\fverb{\setbox\fverbbox=\hbox\bgroup\verb}
\newcommand\fverbdo{\egroup\medskip\noindent%
            \fbox{\unhbox\fverbbox}\ }
\newcommand\fverbit{\egroup\item[\fbox{\unhbox\fverbbox}]}
\newbox\fverbbox

\numberwithin{equation}{section}

\long\def\symbolfootnote[#1]#2{\begingroup%
\def\thefootnote{\fnsymbol{footnote}}\footnote[#1]{#2}\endgroup}

\newcommand{\be}{\begin{equation}}
\newcommand{\ee}{\end{equation}}

\newcommand{\bea}{\begin{eqnarray}\begin{aligned}}
\newcommand{\eea}{\end{aligned}\end{eqnarray}}
\newcommand{\mat}{\begin{pmatrix}}
\newcommand{\rix}{\end{pmatrix}}

\renewcommand{\bar}{\overline}

\newcommand{\go}{{\tilde g}}

\newcommand{\q}{\quad}
\newcommand{\qq}{\qquad}
\newcommand{\qqq}{\qquad\quad}

\newcommand{\beqa}{\begin{eqnarray}}
\newcommand{\eeqa}{\end{eqnarray}}
\newcommand{\lp}{\left(}
\newcommand{\rp}{\right)}

\newcommand{\beq}{\begin{equation}}
\newcommand{\eeq}{\end{equation}}
\newcommand{\hc}{\mbox{h.c.}}

\newcommand{\abs}[1]{\left\vert#1\right\vert}

\newcommand{\order}[1]{{\cal O}\left(#1\right)}

\newcommand{\CM}{\mathcal{M}}

\newcommand{\chifv}{\large$\chi$\normalsize FV}
\newcommand{\FormFlavor}{$\mathtt{FormFlavor}$}

\begin{document}

\title{Chiral Flavor Violation\\ from \\ Extended Gauge Mediation}

\date{\today}

\author[1]{Jared A.~Evans}
\author[2]{David Shih}
\author[2]{Arun Thalapillil}

\affil[1]{\small{Department  of Physics\\ 
University of Illinois at Urbana-Champaign\\ 
Urbana, IL 61801}}
\affil[2]{NHETC\\ 
Department of Physics and Astronomy\\
Rutgers University\\
Piscataway, NJ 08854 }

\maketitle

\begin{abstract}
Models of extended gauge mediation, in which large $A$-terms arise through direct messenger-MSSM superpotential couplings, are well-motivated by the discovery of the 125~GeV Higgs. However, since these models are not necessarily MFV, the flavor constraints could be stringent. In this paper, we perform the first detailed and quantitative study of the flavor violation in these models. To facilitate our study, we introduce a new tool called \FormFlavor\ for computing precision flavor observables in the general MSSM. We validate \FormFlavor\ and our qualitative understanding of the flavor violation in these models by comparing against analytical expressions.  Despite being non-MFV, we show that these models are protected against the strongest constraints by a special flavor texture, which we dub chiral flavor violation ($\chi$FV). This  results in only mild bounds from current experiments, and exciting prospects for experiments in the near future.
\end{abstract}

\newpage

\tableofcontents

\newpage

\section{Introduction}
\label{sec:intro}

The SUSY flavor problem is a serious challenge for models of weak-scale supersymmetry. (For a review and original references, see \cite{Martin:1997ns,Isidori:2010kg}.) Soft SUSY-breaking introduces many sources of flavor violation beyond the Standard Model Yukawas. Generic points of the MSSM parameter space are ruled out by myriad precision flavor constraints, such as neutral meson mixing and $b\to s\gamma$. Clearly, the underlying mechanism that generates these soft SUSY-breaking parameters from the hidden sector must be quite special. Historically, the lore has been that the mediation of SUSY-breaking must either be flavor-blind or obey the minimal flavor-violating (MFV) ansatz.  Gauge mediated SUSY breaking (GMSB), which is manifestly flavor-blind, is one of the simplest solutions to the SUSY flavor problem (see \cite{Giudice:1998bp} for a review and original references). 

The 2012 discovery of a Standard Model-like Higgs with a mass near $125$ GeV~\cite{Aad:2012tfa,Chatrchyan:2012ufa} presents interesting challenges for models of GMSB, especially in the MSSM, where a 125~GeV Higgs implies either very heavy stops $\gtrsim 10$~TeV or large (multi-TeV) stop $A$-terms \cite{Hall:2011aa, Heinemeyer:2011aa,Arbey:2011ab,Draper:2011aa,Carena:2011aa}. The heavy stop scenario is more fine-tuned and less interesting from both an experimental and theoretical point of view. The large $A$-term scenario allows for light stops, but a mechanism is required to generate these $A$-terms, which are absent at the messenger scale in GMSB.  

Large $A$-terms can arise if the usual GMSB framework is extended to also include direct MSSM-messenger superpotential couplings.  Using light ($\sim1$ TeV) stops and large $A$-terms, these models of extended GMSB (EGMSB) can give rise to the observed Higgs mass at fine-tuning levels close to the best achievable within the MSSM  \cite{Evans:2012hg,Kang:2012ra,
Craig:2012xp,Albaid:2012qk,Abdullah:2012tq,Perez:2012mj,Kim:2012vz,Byakti:2013ti,Craig:2013wga,Evans:2013kxa,Calibbi:2013mka,Fischler:2013tva,Mummidi:2013hba,Knapen:2013zla,Ding:2013pya,Liu:2013vaa,Calibbi:2014pza,Basirnia:2015vga}.  However, since these MSSM-messenger superpotentials are typically flavorful, they are in danger of reintroducing the SUSY flavor problem.  Previous works on EGMSB have either assumed perfect alignment with the third-generation (to get a large stop $A$-term), or considered additional model building (such as Froggatt-Nielsen mechanisms) to ensure this alignment \cite{Chacko:2001km,Shadmi:2011hs,Evans:2011bea,Byakti:2013ti,Evans:2013kxa,Calibbi:2013mka,Galon:2013jba,Jelinski:2014uba,Calibbi:2014yha}. In this paper, we will not presume any such alignment, and we will instead perform the first comprehensive study of the general flavor constraints on EGMSB models for the Higgs mass.  

The precursor to this work was the complete classification of all renormalizable EGMSB couplings consistent with perturbative $SU(5)$ unification provided in \cite{Evans:2013kxa} (see also \cite{Byakti:2013ti}).  By turning on one coupling at a time (perfect alignment with the third generation was assumed) and imposing the Higgs mass constraint, the landscape of EGMSB models was surveyed for their phenomenology and fine-tuning.  It was shown that the EGMSB models that exhibit the smallest tuning are of the form,
\beq
\label{eq:bestModels}
W \ni \kappa_3 Q_3 \Phi \tilde \Phi, \qq W \ni \kappa_3 \bar U_3 \Phi \tilde \Phi, \qq W \ni \kappa_3 Q_3 H_u \Phi, \q \mbox{or} \q W \ni \kappa_3 \bar U_3 H_u \Phi.
\eeq
Here $\Phi$, $\tilde\Phi$ are messenger fields transforming in appropriate representations of the SM gauge group. The first two (second two) models involve a single MSSM field (two MSSM fields) and so they were classified as ``type I" (``type II") models. 

Notice that the least fine-tuned models in (\ref{eq:bestModels}) are all flavorful. Type I Higgs models, i.e., $W \ni \kappa H_u \Phi \tilde \Phi$, while MFV, are more finely-tuned due to the ``little $A/m_H^2$ problem" \cite{Craig:2012xp}.  These models have an irreducible contribution to $m_{H_u}^2=A_{\tilde t}^2+\dots$ where $A_{\tilde t}$ is the stop $A$-term; thus a large $A_{\tilde t}$ is tied to a large $m^2_{H_u}$ and one does not improve tuning in these models by increasing the $A$-terms.

 In this paper, we will study the effects that arise when the couplings of (\ref{eq:bestModels}) are no longer required to align perfectly with the third-generation. For simplicity, we will specialize to a pair of representative type I $Q$-class and $U$-class models, the models I.9 and I.13 from \cite{Evans:2013kxa},
\beq
\label{eq:bestModelsFV}
W \ni \kappa_i Q_i \Phi_D \Phi_L, \qq W \ni \kappa_i \bar U_i \Phi_{D_1} \Phi_{D_2}
\eeq
where the messengers $\Phi_{D}$ and $\Phi_L$ have the SM gauge quantum numbers of the $D$ and $L$ matter fields. 
We have verified explicitly that the other type I squark models are very similar, both qualitatively and quantitatively.  For reasons we will explain shortly, we expect the type II models are also qualitatively similar in their features. We will not consider the effect of turning on multiple EGMSB couplings, or couplings involving lepton flavor violation. Such couplings can be forbidden by appropriate choices of discrete symmetries. Finally, to focus exclusively on the SUSY flavor problem, we will not consider CP violation in this work, i.e., all the couplings $\kappa_i$ are taken to be real in the mass basis of the standard model particles.  The (possibly stringent) constraints from CP violating observables, such as $\epsilon_K$ and the neutron EDM, will be studied in a forthcoming publication \cite{CPVpaper}.

The interactions in (\ref{eq:bestModelsFV}) result in flavor-violating contributions to the squark mass matrices
\beq
\label{squarkmassmatrices}
  \delta\CM_{\tilde u}^2 =  \lp
\begin{tabular}{c|c}
$  \delta m_Q^2$ &  $s_\beta v A_{\tilde u}^\dagger  $\\
\hline
 $ s_\beta  v A_{\tilde u}$ & $ \delta m_U^2 $ \end{tabular}\rp,
 \qquad \delta\CM_{\tilde d}^2 =  \lp
\begin{tabular}{c|c}
$ \delta m_Q^2$ &  $c_\beta v A_{\tilde d}^\dagger$\\
\hline
 $c_\beta v A_{\tilde d}$ & $ \delta m_D^2$ \end{tabular}\rp
\eeq
  Here each block (clockwise from upper left: $LL$, $LR$, $RR$, $RL$) is a $3\times 3$ matrix.  These should be added to the flavor-conserving GMSB contributions and the supersymmetric contributions, and together they contribute to precision flavor observables through a variety of one-loop diagrams involving the squarks and other superpartners. The calculation of these one-loop diagrams for general MSSM spectra is a tedious and laborious task, ideally suited for a computer program.  There are several such programs that are publicly available; unfortunately, we found that they were all unsuitable for our purpose.  These programs either assumed MFV, did not have a sufficiently broad list of flavor observables, were numerically unstable, or were found to have bugs, likely introduced in transcribing formulas by hand from the literature. 

As a result, we found it necessary to develop a new tool called \FormFlavor\ for the study of flavor physics.  \FormFlavor\ takes general MSSM spectra and computes the contributions to the various flavor observables shown in 
table \ref{tab:obs}. The novel aspect of \FormFlavor\ is that the computation of one-loop Wilson coefficients in the MSSM is done {\it completely from scratch}, using the general-purpose packages $\mathtt{FeynArts}$~\cite{Hahn:2000kx} and $\mathtt{FormCalc}$~\cite{Hahn:1998yk}.\footnote{While this work was in preparation, a new tool $\mathtt{FlavorKit}$~\cite{Porod:2014xia} was published with a very similar approach to calculating flavor observables from scratch. $\mathtt{FlavorKit}$ aims to be even more general, in that it can derive the one-loop Wilson coefficients for a general model, not just the MSSM. It would be interesting to compare the two codes in detail.} This avoids the problems associated with transcribing formulas from the literature, and it facilitates the inclusion of additional flavor observables in an automated and modular way. We intend to make \FormFlavor\ publicly available; its usage and validation will be described in an upcoming publication \cite{FFpaper}.

\begin{table}[tdp]
\begin{center}
\begin{tabular}{|c|c|c|}
\hline
Observable & Experiment & SM prediction \\
\hline 
$\Delta m_{K}$ &$ (3.484\pm0.006) \times 10^{-15}~\rm{GeV}~$ & $-$  \\
\hline
$\Delta m_{B_d}$ &$ (3.36\pm0.02) \times 10^{-13}~\rm{GeV}$ & $ (3.56\pm0.60) \times 10^{-13}~\rm{GeV}$~\cite{Lenz:2014jya}  \\
\hline
$\Delta m_{B_s}$ &$ (1.169\pm0.0014) \times 10^{-11}~\rm{GeV}$ & $ (1.13\pm0.17) \times 10^{-11}~\rm{GeV}$~\cite{Lenz:2014jya} \\
\hline
$\Delta m_{D}$ &$ (6.2^{+2.7}_{-2.8}) \times 10^{-15}~\rm{GeV}$ & $-$ \\
\hline
 ${ Br}(K^+\to \pi^+ \nu \bar{\nu})$ &$   (1.7\pm1.1) \times 10^{-10} $ & $(7.8\pm0.8)\times 10^{-11}$ ~\cite{Brod:2010hi} \\
\hline
${ Br}(B \to X_s \gamma) $ & $ (3.40\pm0.21)\times10^{-4}$ & $(3.15\pm0.23)\times10^{-4}$~\cite{Misiak:2006ab}   \\
\hline
${ Br}(B \to X_d \gamma) $ & $ (1.41\pm0.57)\times10^{-5}$ \cite{delAmoSanchez:2010ae,Crivellin:2011ba} & $(1.54^{+0.26}_{-0.31})\times10^{-5}$~\cite{Crivellin:2011ba}   \\
\hline
${ Br}(B_s \to \mu^+ \mu^-)$ &$(2.9\pm0.7)\times 10^{-9}$~\cite{CMSandLHCbCollaborations:2013pla} & $(3.65\pm 0.23)\times 10^{-9}$~\cite{Bobeth:2013uxa}   \\
\hline
${ Br}(B_d \to \mu^+ \mu^-)$ &$ (3.6^{+1.6}_{-1.4}) \times 10^{-10}$~\cite{CMSandLHCbCollaborations:2013pla} & $(1.06\pm 0.09)\times 10^{-10}$~\cite{Bobeth:2013uxa}   \\
\hline
\end{tabular}
\end{center}
\caption{Current experimental values based on PDG and HFAG fits~\cite{{Beringer:1900zz},{Amhis:2012bh}} except where noted.  No reliable theoretical prediction for $\Delta m_D$ currently exists.  Although literature on the subject exists, we do not use a theoretical prediction for $\Delta m_K$ (see the discussion of $\Delta m_K$ in section~\ref{sec:future} for more details). }
\label{tab:obs}
\end{table}

Starting from the points of reduced tuning identified in \cite{Evans:2013kxa}, we will turn on $\kappa_{1,2}$ and use \FormFlavor\ to  investigate the constraints from precision flavor observables. Given that they are not MFV, one might expect these constraints to be extremely stringent. Generic operator bounds put the scale of flavor violation at $20$ PeV or higher from $K-\bar K$ and $D-\bar D$ mixing  \cite{Isidori:2010kg}. Even in SUSY, where one benefits from loop factors, etc., the bounds are close to 500 TeV (without using CP violation) \cite{Altmannshofer:2009ne, Altmannshofer:2013lfa}. However,  we will find that in these EGMSB models, the limits from flavor-violation are extremely mild -- to the point that $\kappa_{1,2,3}$ can all be the same size and yet the model is not ruled out by flavor!

Much of this paper  will be devoted to identifying the reasons for these surprisingly mild flavor constraints. Although the overall heavier mass scale required to raise the Higgs mass plays a role, the most important reason is the fact that the EGMSB models (\ref{eq:bestModels}) only violate flavor through a spurion of either $SU(3)_Q$ or $SU(3)_U$, but not both.  Because of this {\it chiral flavor violation} (\chifv), EGMSB models have a novel flavor texture -- flavor violation primarily occurs only in either the left-chiral sector of the squark mass-squared matrix (for the $Q$-class models) or the right-chiral sector of the squark mass-squared matrix (for the $U$-class models). Only via communication through the SM Yukawas can the other chiral sector feel the flavor violation. As it turns out,  the most stringent flavor constraints, which come from $\Delta m_{K}$ and $\Delta m_{D}$, are vastly reduced when the flavor violation is restricted to only the left- or right-chiral sector.  (It is also on these general grounds that we expect the type II models are similarly unconstrained by flavor, although it would be interesting to verify this in detail.) In a forthcoming work \cite{ChiFVpaper}, we will study this new ansatz of \chifv\ in more generality, along the lines of what has been done for MFV.  \chifv\ represents an interesting intermediate case between full flavor anarchy (which is known to be heavily constrained) and MFV (which is known to be basically unconstrained). 
  
In order to validate and interpret our numerical findings, we will compare them against analytic expressions for the flavor observables. Historically, the mass insertion approximation (MIA) has been utilized to interpret the influence of flavor-violating squark masses on precision flavor observables (see e.g.,  \cite{Isidori:2010kg} for a review of the MIA and original references).  However, the utility of the MIA is limited when one or more of these mass insertions are $\order{1}$. Since that is precisely the interesting region of parameter space for our EGMSB models, the traditional MIA cannot be used here.
Fortunately, in these EGMSB models there is another handle we can use to obtain an analytic understanding.  From two powers of our anti-fundamental spurion, $\kappa_a$, we can construct an adjoint+singlet of $SU(3)_Q$ or $SU(3)_U$,
\beq
K_{ab} \equiv \kappa_{a}^*\kappa_{b} 
\label{eq:notation}
\eeq 
This matrix $K$ governs all flavor violation from EGMSB. By exploiting the fact that it is only rank 1,  together with the special properties of the soft masses for the type I models, we are able to obtain analytic formulas for the flavor observables that treat the flavor violation {\it exactly}. Then the only expansion that we do is in $v/m_{SUSY}$, which is still an excellent approximation in these models, even when the flavor violation is $\CO(1)$. With this technique, we are able to make precise estimates of the supersymmetric contributions to flavor observables, validate our numerical results in detail, and understand their qualitative features.  

Our paper is outlined as follows. In section~\ref{sec:preliminaries}, we describe the idea behind chiral flavor violation and how it applies to our models (\ref{eq:bestModels}) in restricting the texture of the squark mass matrices.  We then further restrict our attention to the type I squark models, and highlight some of their special features that will be useful in analyzing the flavor observables. In section~\ref{sec:flavor}, we turn to a detailed study of the flavor constraints on the EGMSB models (\ref{eq:bestModelsFV}). We illustrate these constraints  in the $\kappa_{1}$-$\kappa_2$ plane using a series of plots of flavor observables computed with \FormFlavor. We also provide an analytic understanding of the features of these plots using the special properties of rank 1 \chifv\ and the type I squark models.  Finally in section~\ref{sec:conclusions}, we conclude with a brief summary of our results, and a discussion of the promising future prospects for precision flavor tests of these EGMSB models. In the appendices, we detail our parameter deformation and various subtleties that arise there, provide a brief description of \FormFlavor\ (postponing a more detailed manual and validation for an upcoming work \cite{FFpaper}), compile the necessary expressions for flavor observables using a uniform notation, and present some other formulas used in this work.

\section{Chiral Flavor Violation and EGMSB}
\label{sec:preliminaries}

\subsection{Chiral Flavor Violation}
\label{sec:chiFV}

In the absence of Yukawa couplings, the Standard Model flavor symmetry group is
\beq
G_f = SU(3)_Q\times SU(3)_U\times SU(3)_D\times SU(3)_L\times SU(3)_E
\eeq
The assumption of MFV is that this flavor symmetry is broken only by the Yukawa couplings $y_u$, $y_d$ and $y_\ell$, which transform as $({\bf 3},{\bf \bar 3})$ under $SU(3)_U\times SU(3)_Q$, $SU(3)_D\times SU(3)_Q$, and $SU(3)_E\times SU(3)_L$, respectively. As described in the introduction, the chiral flavor violation (\chifv) ansatz goes beyond MFV, and postulates that, in addition to the Yukawas, the flavor symmetry is broken by additional spurions transforming only under a single $SU(3)$ of the full flavor group. 

The \chifv\ ansatz results in a special texture of flavor violation in the MSSM, one which greatly suppresses the constraints from many flavor violating observables.  In general, the $3\times 3$ soft mass-squareds $m_{X}^2$ transform as an adjoint+singlet of $SU(3)_X$, while the $A$-terms transform in the same way as the Yukawas. Using spurions from only a single $SU(3)_{X}$, one can obtain flavor violation only in $m_{X}^2$. To obtain flavor violation in any of the other soft terms, one must involve the Yukawa couplings. Thus these other soft masses inherit an MFV-like suppression.

The cases of interest for this paper are when $X=Q$ or $U$. (We reserve a more general treatment for an upcoming publication \cite{ChiFVpaper}.) 
Let us now examine these in more detail.   To leading order in the Yukawa couplings, the symmetries of $Q$-class \chifv\, imply  that 
\beq\label{SigmaGammadefQ}
\delta m_U^2= y_u \Sigma y_u^\dagger,\qquad \delta m_D^2= y_d \Sigma y_d^\dagger,\qquad  A_{\tilde u}=  y_u\Gamma,\qquad  A_{\tilde d}=  y_d\Gamma
\eeq
where $\Sigma$ and $\Gamma$ are built out of the $SU(3)_Q$ spurions. Substituting these into (\ref{squarkmassmatrices}), we have under the third-generation dominant approximation (whereby $y_u$ and $y_d$ are nonzero only in the 33 component): 
\bea
\label{eq:Qtexturegenu}
& \delta \CM_{\tilde u}^2 \approx \lp \begin{tabular}{c|c}
$\delta m_Q^2$ &  \begin{tabular}{ccc} $0$ & $0$ & $m_t \Gamma_{31}^*$ \\ $0$ & $0$ & $m_t \Gamma_{32}^*$\\ $0$ & $0$ & $m_t \Gamma_{33}^*$\end{tabular}\\
\hline
 \begin{tabular}{ccc} $0$ & $0$ & $0$ \\ $0$ & $0$ & $0$ \\ $m_t \Gamma_{31} $ & $m_t \Gamma_{32} $ & $m_t \Gamma_{33}$\end{tabular} & \begin{tabular}{ccc} $0$ & $0$ & $0$\\ $0$ & $0$ & $0$ \\ $0$ & $0$ & $y_t^2\Sigma_{33}$ \end{tabular} \end{tabular}\rp\\
 & \delta \CM_{\tilde d}^2 \approx \lp \begin{tabular}{c|c}
$\delta m_Q^2$ &  \begin{tabular}{ccc} $0$ & $0$ & $m_b \Gamma_{31}^*$ \\ $0$ & $0$ & $m_b \Gamma_{32}^*$\\ $0$ & $0$ & $m_b \Gamma_{33}^*$\end{tabular}\\
\hline
 \begin{tabular}{ccc} $0$ & $0$ & $0$ \\ $0$ & $0$ & $0$ \\ $m_b \Gamma_{31} $ & $m_b \Gamma_{32} $ & $m_b \Gamma_{33}$\end{tabular} & \begin{tabular}{ccc} $0$ & $0$ & $0$\\ $0$ & $0$ & $0$ \\ $0$ & $0$ & $y_b^2\Sigma_{33}$ \end{tabular} \end{tabular}\rp
 \eea
We see that within the third-generation dominant approximation, no flavor violation appears in the RR block for either the up or down squark mass matrices. Also, the only flavor violation in the LR block involves the 3rd generation, and is $v/m_{SUSY}$ suppressed. These features greatly reduce the sensitivity of $Q$-class \chifv\ to precision flavor constraints. As we will explain in more detail in section \ref{subsubsec:mesonmixing}, what typically provide the most stringent flavor bounds on new physics (i.e., $\Delta m_{K}$ and $\Delta m_D$) involve the 1st and 2nd generations and involve simultaneous violation of flavor in both the left \emph{and} right chiral sectors.

Meanwhile the $U$-class \chifv\ models are even more insulated from constraints. Here, on symmetry grounds, and again to leading order in the Yukawas, we must have 
\beq\label{SigmaGammadefU}
\delta m_Q^2=y_u^\dagger \Sigma y_u,\qquad A_{\tilde u}= \Gamma y_u,\qquad \delta m_D^2=A_{\tilde d}=0
\eeq
where now $\Sigma$ and $\Gamma$ are built out of the $SU(3)_U$ spurions.  Again substituting these into (\ref{squarkmassmatrices}) in the third-generation dominant approximation, we find:
\beq
\label{eq:Utexturegenu}
\delta \CM_{\tilde u}^2 
  \approx \lp \begin{tabular}{c|c}
 \begin{tabular}{ccc} $0$ & $0$ & $0$\\ $0$ & $0$ & $0$ \\ $0$ & $0$ & $y_t^2\Sigma_{33}$ \end{tabular} &   \begin{tabular}{ccc} $0$ & $0$ & $0$ \\ $0$ & $0$ & $0$ \\ $m_t \Gamma_{13}^* $ & $m_t \Gamma_{23}^* $ & $m_t \Gamma_{33}^*$\end{tabular}\\
\hline
 \begin{tabular}{ccc} $0$ & $0$ & $m_t \Gamma_{13}$ \\ $0$ & $0$ & $m_t \Gamma_{23}$\\ $0$ & $0$ & $m_t \Gamma_{33}$\end{tabular} & $\delta m_U^2$ \end{tabular}\rp,\quad \delta \CM_{\tilde d}^2  \approx \lp \begin{tabular}{c|c}
 \begin{tabular}{ccc} $0$ & $0$ & $0$\\ $0$ & $0$ & $0$ \\ $0$ & $0$ & $y_t^2 \Sigma_{33}$ \end{tabular}  & 0 \\ \hline 0  & 0 \end{tabular}\rp
 \eeq
Thus there is no down-type flavor violation at all in the third-generation dominant approximation! As the majority of sensitive flavor probes involve the down sector, this eliminates almost all contributions to flavor observables. The only exception is $\Delta m_D$, which again has reduced sensitivity because there is no simultaneous left-right flavor violation in the 1st/2nd generations.   (Down-type flavor observables can involve the up-squark RR block through chargino loops, but these must be mostly Higgsino-like, so they will be suppressed by Yukawa couplings.) 

Finally, we should comment on the role of the RG. The RGEs from the messenger scale to the weak scale add more terms, but the symmetry-based arguments given above -- which were truncated at leading order in the Yukawas -- clearly continue to hold with the inclusion of higher orders in the Yukawas.  In particular, terms that are zero in the third-generation dominant limit remain zero under RG evolution.   For this reason, we do not need to concern ourselves with the details of RG running to gain a qualitative understanding of the EGMSB flavor-violating contributions to our flavor observables.

\subsection{$\chi$FV in Type I Squark Models}
\label{rank1miasec}

The EGMSB models (\ref{eq:bestModels}) are clearly examples of $Q$-class or $U$-class \chifv. In this subsection, we will discuss some further features that are specific to the type I squark models that will help us analyze the Wilson coefficients for flavor observables in the next section.

We begin by quoting the explicit formulas for the soft masses for general type I squark models (see appendix \ref{app:softTypeI}).  For concreteness, we focus on $Q$-class models, where we have (in addition to the GMSB contributions),
\bea
\label{eq:QgenK}
& \delta m_{Q}^2 = { d_Q \over 256\pi^4}\Bigg( (  d_\phi + d_Q) \kappa^2 - 2 C_r g_r^2  - \frac{16\pi^2}{3} h\! \lp\!\frac{\Lambda}{M}\!\rp \!\! \frac{\Lambda^2}{M^2} \Bigg)K\Lambda^2 \\
& \delta m_{U}^2 =- {d_Q d^{QH}_U \over 256\pi^4}  y_u K y_u^\dagger   \Lambda^2\\
& \delta m_{H_u}^2 = - {3 d_Q \over 256\pi^4} \Tr[ y_u K  y_u^\dagger] \Lambda^2 \\
&A_{\tilde u} =-{d_Q \Lambda  \over16\pi^2} y_u K .
\eea
Here we are neglecting the (numerically irrelevant) down-Yukawa contributions;  $K$ is the rank one matrix of couplings defined in (\ref{eq:notation}); $d_U^{QH}=2$ is a multiplicity factor; and $h(x)$ is an $\order{1}$ loop function (see (\ref{eq:hfunc}) for  the exact form).  For our $Q$-class model of study (\ref{eq:bestModelsFV}),  we have $d_Q= N=6$, $d_\phi=5$, where $N$ is the number of messengers.\footnote{For the similar $U$-class model formulas, simply change $U\leftrightarrow Q$, $A_u\to A_u^\dagger$, $y_q\to y_q^\dagger$, and $K\to K^T$.  The analogous multiplicity factor is $d_Q^{UH}=1$, and for our $U$-class model, we have $d_U=2N=6$, $d_\phi=4$.}

We can see the \chifv\ texture of (\ref{eq:Qtexturegenu}) quite clearly in the formulas for $\delta m_U^2$ and $A_{\tilde u}$ in  (\ref{eq:QgenK}).   There are also additional features of these explicit formulas that go beyond the \chifv\ ansatz. We notice that $\delta m_Q^2$, together with the quantities $\Gamma$ and $\Sigma$ introduced in (\ref{SigmaGammadefQ}), are all proportional to $K$:
\beq
\label{KpropA}
\delta m_Q^2 = \beta K,\qquad \Gamma = \gamma K ,\qquad \Sigma=\sigma K
\eeq
The simplicity of these relations is partly  due to the rank 1 nature of the flavor violation. But in principle, on symmetry grounds alone, there could have been additional contributions to (\ref{KpropA}) proportional to the identity matrix and powers of the Yukawa couplings. These are absent due to the specific form of the type I squark couplings.  This is the main reason we have focused on the type I models in this paper. The forms of the soft masses are more complicated in the type II models, and while we expect them to be similarly protected by their \chifv\ flavor texture, understanding them at the level of analytical detail that we apply to the type I models is more difficult. 

Using the relations (\ref{KpropA}), we now discuss the diagonalization of the squark mass matrices. Because the LR blocks are suppressed by $v/m_{SUSY}$,  the squark mass eigenvalues are given by those of the LL and RR blocks to a very good approximation. The RR blocks are already diagonal  in the third-generation dominant approximation according to  (\ref{eq:Qtexturegenu}). The down-squark RR masses are just their minimal GMSB values $m_0^2$, while the up-squark RR masses are given by $m_0^2$, $m_0^2$ and $m_0^2 +\sigma y_t^2 \kappa_3^2\equiv m_{RR}^2$. In practice the third eigenvalue is only a little offset 
from $m_0^2$ throughout the parameter space of our model, and  this difference can be neglected. 

The situation for the LL block is only a little more complicated. Since $\delta m_Q^2$ is just proportional to $K$, the LL block of the squark mass matrices is diagonalized by the unitary transformation
\beq
\label{ULL}
U_{LL} = \left( \begin{matrix} \hat\kappa_1 & a_1 & b_1 \\ \hat\kappa_2 & a_2 & b_2 \\ \hat\kappa_3 & a_3 & b_3 \end{matrix}\right)
\eeq
where $\hat \kappa_a=\kappa_a/\kappa$ and $\vec a$ and $\vec b$ are any two orthonormal basis vectors for the orthogonal subspace $\vec\kappa^{\perp}$. The eigenvalues are $m_0^2$, $m_0^2$, and
\begin{equation}
\label{mSdef}
m_{S}^2\equiv m_0^2 + \beta \kappa^2
\end{equation}
(Here we are ignoring the small differences between the minimal GMSB contributions to the $Q$, $U$, $D$ soft masses coming from $SU(2)\times U(1)$.) In our EGMSB models, $m_{S}^2$ corresponds roughly to the mass of the lightest squark, and moreover
\beq\label{mSllm0}
m_S^2\ll m_0^2
\eeq
Thus the Wilson coefficients will generally be dominated by this lightest squark running in the loop, and  $m_S^2$ will play a central role in controlling the size of the flavor-violating effects in these models. In fact, because of (\ref{mSllm0}), diagrams where the right-handed squarks propagate tend to be suppressed, and we will see that it suffices to focus on the LL flavor violation exclusively.

\section{Flavor Observables and Constraints on EGMSB}
\label{sec:flavor}

\begin{figure}[!t]
\begin{center}
\includegraphics[scale=0.82]{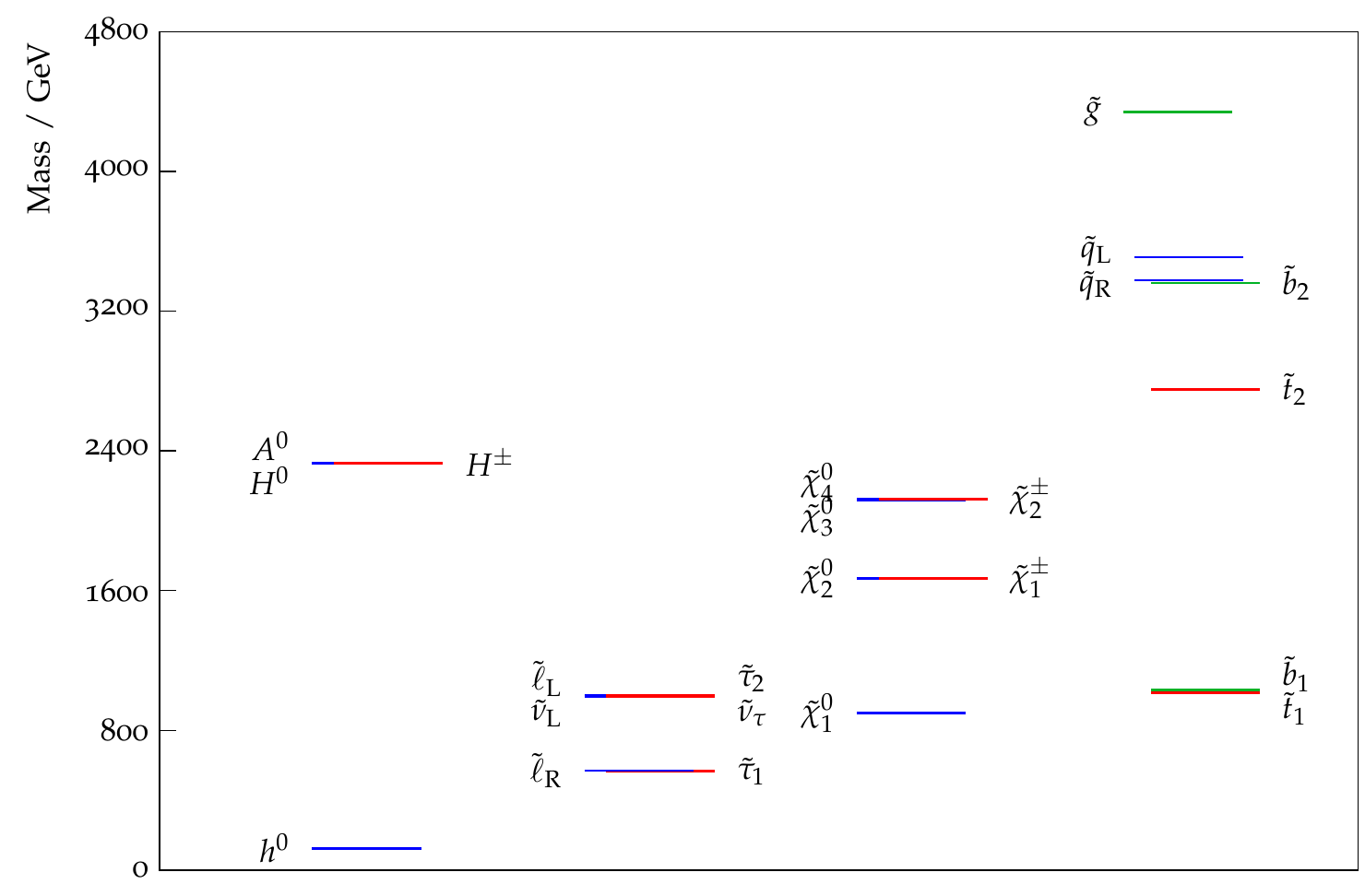}
\includegraphics[scale=0.82]{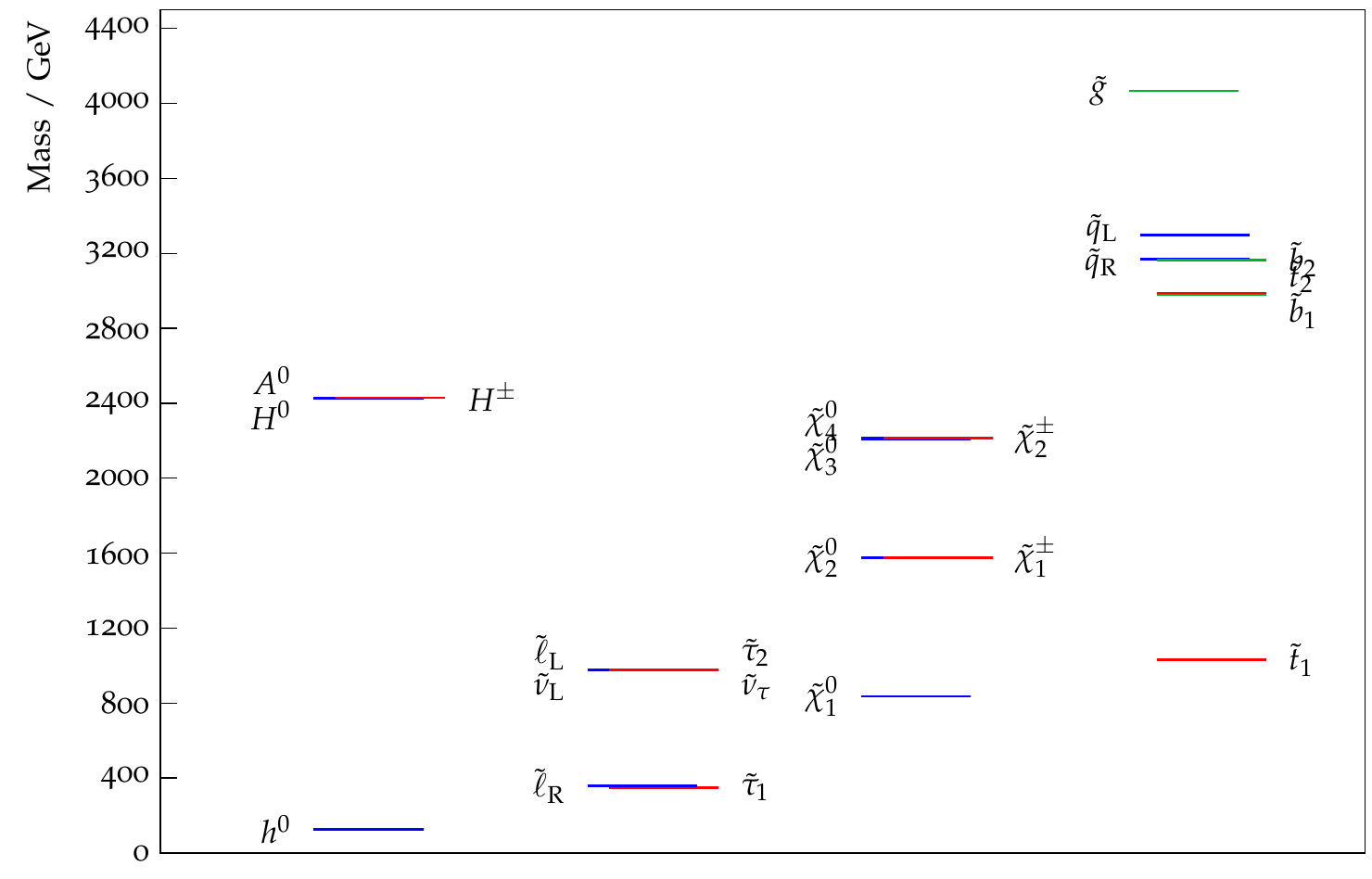}
\end{center}
\caption{{\bf Top:} Mass spectrum of the $\kappa_i Q_i \Phi_D \Phi_L$ model at $\kappa_1=\kappa_2=0$. {\bf Bottom:} Mass spectrum of the $\kappa_i U_i \Phi_{D_1} \Phi_{D_2}$ model at $\kappa_1=\kappa_2=0$.}
\label{fig:spectra}
\end{figure}

In this section, we investigate the flavor constraints on our  EGMSB models (\ref{eq:bestModelsFV}) in detail. For reference, the spectra corresponding to $\kappa_1=\kappa_2=0$ are shown in fig.~\ref{fig:spectra}. These are essentially the same points that were identified in \cite{Evans:2013kxa} as being the least fine-tuned EGMSB models with $m_h=125$~GeV.   Starting from these flavor-aligned points, we will perform a numerical scan in the flavor-violating parameter space ($\kappa_1/\kappa_3$, $\kappa_2/\kappa_3$) of the models, using  \FormFlavor\ to compute the flavor observables and compare against their experimental values. Going through each observable $X$ from table~\ref{tab:obs} in turn, we will exhibit plots of
\bea
{ |\left[X\right]_{\text{\tiny{TH}}}-\left[X\right]_{\text{\tiny{EXP}}}|\over \left[\sigma(X)\right]_{\text{\tiny{TH+EXP}}}}
\eea
Here $\left[X\right]_{\text{\tiny{TH}}}$ denotes the SM prediction together with the EGMSB contribution, and $\left[\sigma(X)\right]_{\text{\tiny{TH+EXP}}}$ is the theoretical and experimental errors added in quadrature. A contour of 2, which is roughly the 95\% CL exclusion limit, will be taken to indicate the point at which the EGMSB model is excluded by the given flavor observable.\footnote{Two exceptions to this are $\Delta m_K$ and $\Delta m_D$ where the theory errors are uncontrolled. In these cases, we will plot instead:
\bea
{ |\left[\Delta m_X\right]_{\text{\tiny{EGMSB}}}|\over \left[\Delta m_X+2\sigma(\Delta m_X)\right]_{\text{\tiny{EXP}}}}
\eea
Thus a contour of 2 does not represent a 95\% confidence level exclusion for these observables. However, as it would require a substantial tuning for the standard model and the new physics contributions to cancel against one another, values larger than 1 are suspect.}

As we deform away from $\kappa_1=\kappa_2=0$, there are numerous subtleties that must be taken into account regarding how the other parameters of the model are varied. These subtleties and specifics of the procedure are described in appendix \ref{app:scan}. In short, $\kappa_1$ and $\kappa_2$ are introduced in such a way that the superpartner mass eigenvalues and ``net" $A$-terms are essentially held fixed.

In order to validate the numerical results from  \FormFlavor, we will compare them against analytical formulas for the flavor observables. This will also shed further qualitative insights on the role of \chifv\ in weakening the flavor constraints. As discussed in the Introduction, the usual mass insertion approximation fails due to $\order{1}$ entries. Instead, we will use flavor symmetries and the special features of the type I models discussed above to characterize their {\it exact} $\kappa$ dependence, to leading order in an expansion in $v/m_S$. In general, the $\kappa$ dependence arises through the lightest squark mass (\ref{mSdef}) and the unitary matrix $U_{LL}$ given in (\ref{ULL}). 
(The LR and RL blocks in (\ref{eq:Qtexturegenu}) also depend on $\kappa$ through (\ref{KpropA}), but as discussed in the previous section, this dependence can be generally be neglected due to the heaviness of the right-handed squarks and the extra $v/m_S$ suppression.) The dependence  on $U_{LL}$ is constrained by the flavor symmetries, and as explained in the previous paragraph, the squark mass eigenvalues are mostly held fixed in our parameter space. Thus it will be possible to fully characterize the features of the  \FormFlavor\ plots in terms of very simple functions of $\kappa$.

\subsection{Q-class Models}


\subsubsection{Meson Mixing}\label{subsubsec:mesonmixing}

We begin with the $\Delta F=2$ meson mixing observables.  As we discuss in appendix \ref{app:Dmx}, for $X=K$ and $D$,
\beq
\Delta m_X = 2\,{\rm Re} \,\langle \bar X|H_{eff}|X\rangle
\eeq
to a good approximation, while for $X=B_d$ and $B_q$,
\beq
\Delta m_X = 2| \langle \bar X|H_{eff}|X\rangle|
\eeq
to a good approximation. Here $H_{eff}$ is the $\Delta F=2$ effective Hamiltonian; its local short-distance part (relevant for the MSSM contributions) is built out of the four-fermi operators,
\bea
\label{fourfermi}
& (\CO_S^{MN})_{ab} = ( \bar q_a   P_M q_b) ( \bar q_a   P_N q_b)\\
& (\CO_V^{MN})_{ab} = ( \bar q_a \gamma^\mu P_M q_b) ( \bar q_a \gamma_\mu P_N q_b)\\
& (\CO_T^{MN})_{ab} = ( \bar q_a \sigma^{\mu \nu}P_M q_b) ( \bar q_a \sigma_{\mu \nu} P_N q_b)
\eea
Here $M,N=L,R$ label the chirality of the incoming quarks; and $ab=12,12,13,23$ for $\Delta m_K$, $\Delta m_D$, $\Delta m_{B_d}$ and $\Delta m_{B_s}$, respectively, while $q$ is up-type for $\Delta m_D$ and down-type for the rest.

\begin{figure}[!t]
\includegraphics[scale=0.75]{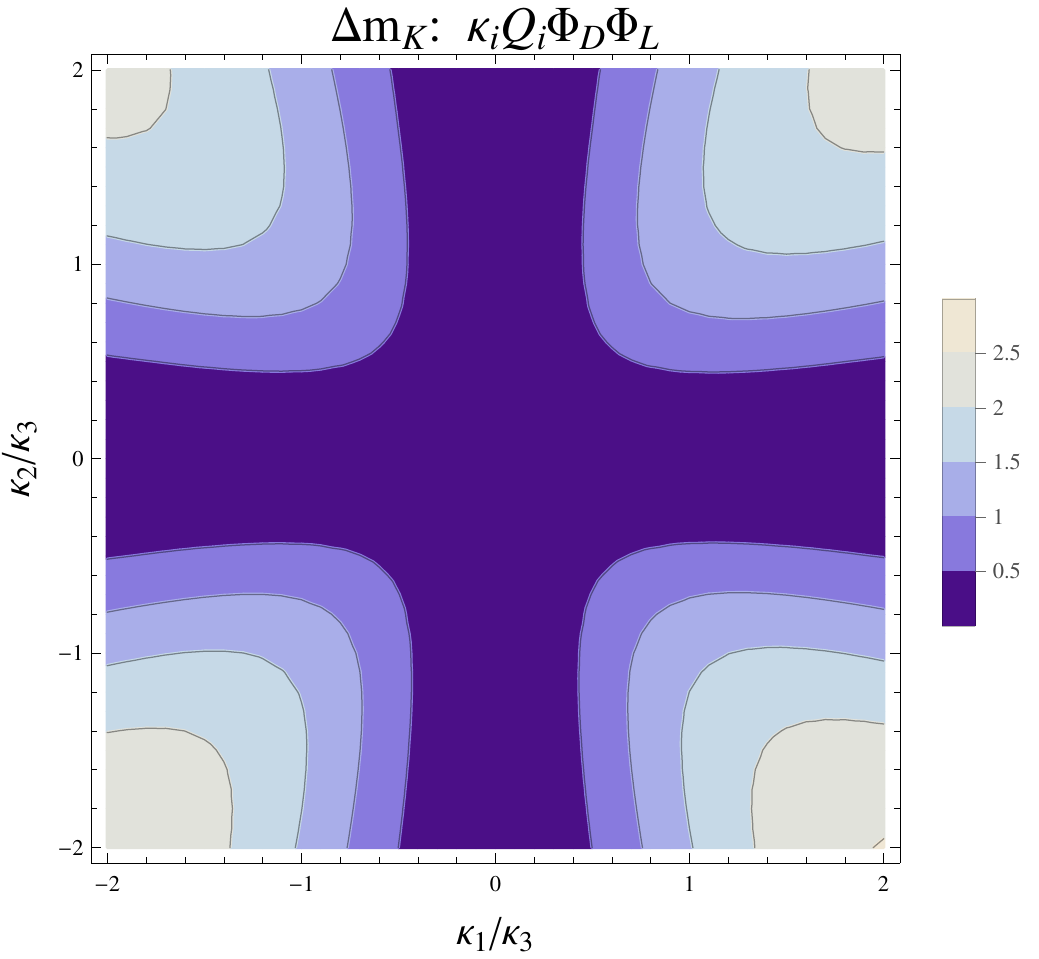}
\includegraphics[scale=0.75]{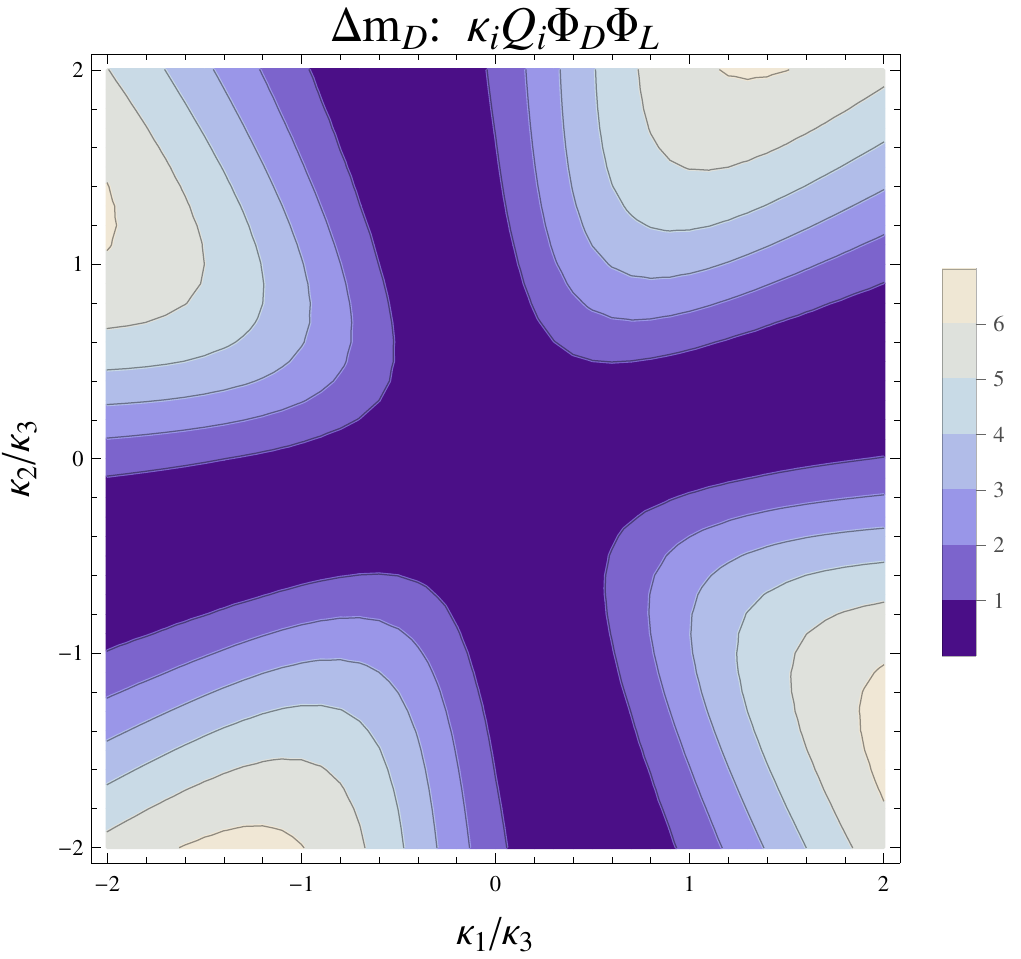}\\
\includegraphics[scale=0.75]{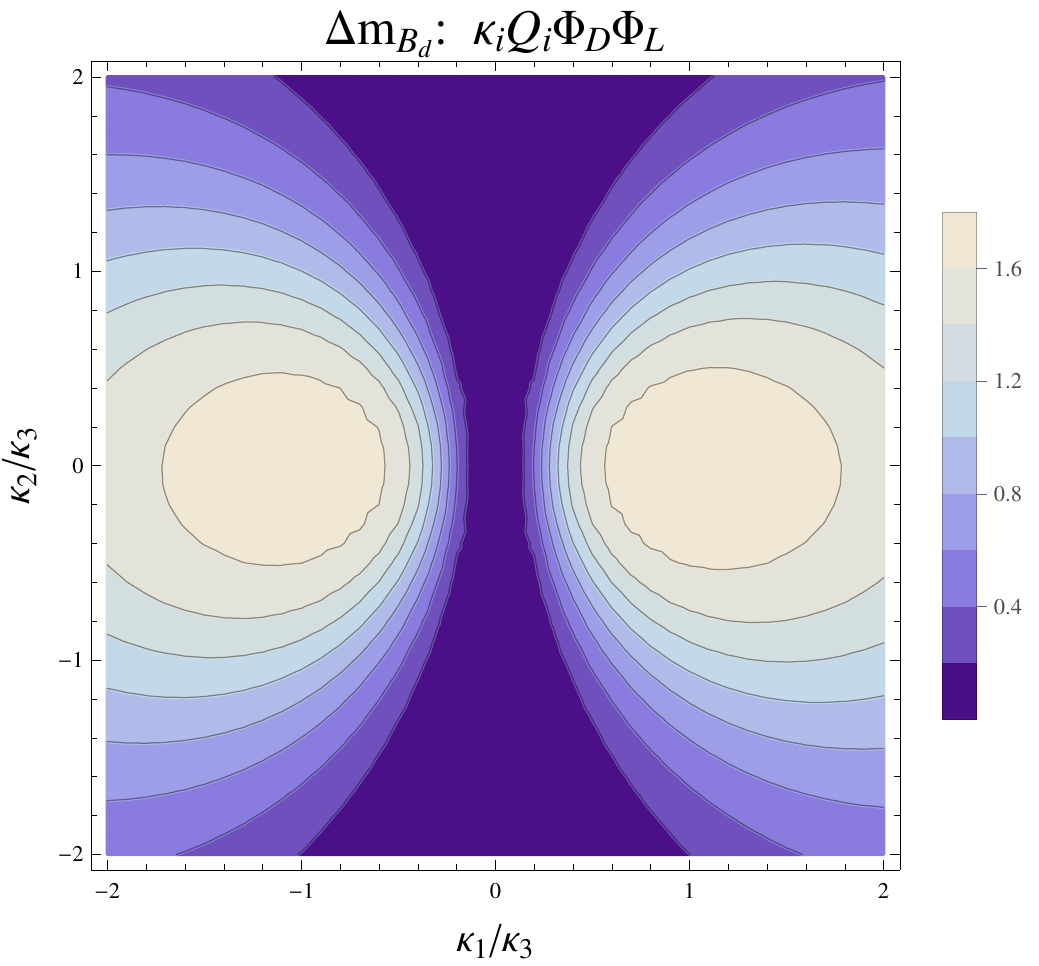}
\includegraphics[scale=0.75]{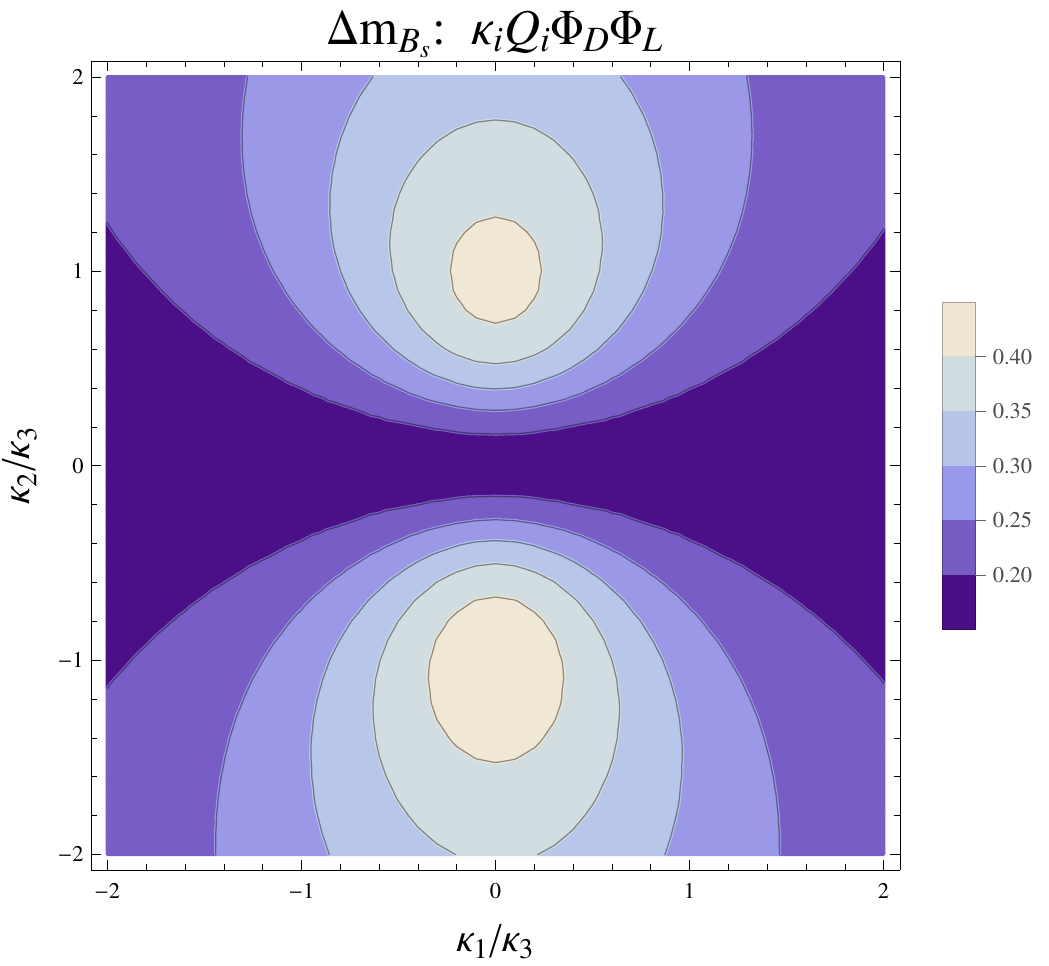}
\caption{Plots of meson mixing observables the $Q$-class model, $\kappa_i Q_i \Phi_D\Phi_L$.  $\Delta m_K$, $\Delta m_D$, $\Delta m_{B_d}$, $\Delta m_{B_s}$ are presented in the upper left, upper right, lower left and lower right, respectively.  Both $\Delta m_K$ and $\Delta m_D$ are presented as the \FormFlavor\ output over the experimental value, while $\Delta m_{B_d}$ and $\Delta m_{B_s}$ are the difference between \FormFlavor\ and the experimental value in units of the uncertainty. }
\label{fig:Dmx}
\end{figure}

The full result of  \FormFlavor\ is shown in figure~\ref{fig:Dmx}. In this subsection, we will endeavor to understand its features analytically using the \chifv\ ansatz and the special features of type I EGMSB identified in section \ref{rank1miasec}.

The MSSM contributions to the  $\Delta F=2$ observables are due to box diagrams involving the squarks and the gauginos. Here the great simplification of \chifv\ is that any operator with an $R$ index must transform non-trivially under $SU(3)_{D}$ or $SU(3)_U$; thus it is suppressed by \chifv\ in our $Q$-class models. Furthermore,  the $\CO^{LL}_S$ and $\CO^{LL}_T$ operators all involve an $SU(2)_L$-breaking chirality flip, so they are dropped in the $v=0$ approximation.  Therefore, the only unsuppressed Wilson coefficient is $C_V^{LL}$. This also happens to be the only contribution to the one-loop SM Wilson coefficients, which proceeds through $W$ exchange.

Since $C_V^{LL}$ transforms in the square of the adjoint+singlet representation of $SU(3)_Q$, with just left-handed squarks running in the loop, the only way it can depend on $\kappa$ is,
\begin{equation}
\label{CVLLmesongen}
(C^{LL}_{V})_{ab} =  {\hat\kappa_a^2\hat \kappa_b^2\over m_{S}^2} f_1(m_{S}^2/m_0^2) ,
\end{equation}
where $f_1$ is a dimensionless loop function depending on the LL squark mass eigenvalues.\footnote{All loop functions here and below will implicitly depend on the masses of the other superpartners running in the loop, e.g., the gaugino masses.} Dimensional analysis fixes the dependence on the masses, and the rest of the dependence must be from the unitary matrix (\ref{ULL}).    Under this simplification, the meson mixing contributions are of the form (see appendix \ref{app:Dmx} for more details and specific values of the parameters), 
\beq
 [\langle \bar X | H_{eff} | X\rangle]_{\text{\tiny{EGMSB}}} \approx \frac{1}{3} m_X f_X^2 B_{V,X}^{LL} (C_{V}^{LL})_{ab},
\eeq
 where $f_X$ is the decay constant for the meson; and $B^{LL}_{V,X}$ is an $\order{1}$ hadronic parameter.  This simple formula suffices to accurately describe the $\Delta F=2$ flavor observables. 

Focusing on gluino boxes for simplicity,\footnote{Chargino and gluino-neutralino boxes are also typically comparable. However, they do not affect the qualitative discussion here.  In fact, because the charginos enter with the opposite sign of the gluino and gluino+neutralino contributions, the gluinos alone provide a better estimate quantitatively than might be expected.} we find from explicit computation that the leading Wilson coefficient is  given by 
 \be
 \label{eq:mesonwilson}
 (C_V^{LL})_{ab}=- \hat\kappa_a^2\,\hat\kappa_b^2\frac{\alpha_s^2}{36 m_{S}^2} f_{\tilde{g}}^{\Delta M,\text{\tiny{box}}}(x_g, x_{q})
 \ee
This clearly agrees with the general form (\ref{CVLLmesongen}). Here, $x _{q}=m_{S}^2/m_{0}^2$ and $x_g=m_{\tilde{g}}^2/m_{0}^2$, and $f_{\tilde{g}}^{\Delta M,\text{\tiny{box}}}(x_g, x_{q})\approx 0.05$ is a loop function that we define in appendix~\ref{app:Loops}.  As discussed earlier, since the relevant masses are roughly held fixed in our deformation, the loop function does not change across the parameter space.   The QCD RG running from the SUSY scale to the meson scale is fairly mild for $C^{LL}_V$,  yielding only a $20-30\%$ suppression in the size of the Wilson coefficient \cite{Buras:2001ra}.

The hadronic factors $f_X^2 B^{LL}_{V,X}$ are fairly similar across the mesons. We can define,
\beq 
H_X\equiv \frac{ f_X^2 B^{LL}_{V,X}}{f_K^2 B^{LL}_{V,K}}; \;\; H_K=1,\; H_D\approx H_{B_d}\approx 2,\; H_{B_s}\approx 3
\eeq
So we expect deviations of the form,
\bea
\label{eq:mesonmxsimp}
 [\langle \bar X | H_{eff} | X\rangle]_{\text{\tiny{EGMSB}}} \sim  -  \hat \kappa_a^2\, \hat \kappa_b^2 H_X m_X  \lp 10^{-13} \rp,
\eea
Noting that $ \hat \kappa_a^2\, \hat \kappa_b^2$ is at most $\frac14$, and comparing against table \ref{tab:obs}, we see that $\Delta m_K$ and $\Delta m_D$ should be most sensitive to EGMSB, while $\Delta m_{B_d}$ should be barely sensitive, and $\Delta m_{B_s}$ completely insensitive.\footnote{As discussed in the Introduction,  we are not considering CP violation in this work, in particular $\epsilon_K$.  Although it depends on the precise value of the CP violating phase, the expectation is that this will place a meaningful, tighter constraint on the parameter space when the phase is large.  This and other CPV observables will be studied in an upcoming paper \cite{CPVpaper}. } 

These sensitivities are observed in the plots shown in figure~\ref{fig:Dmx}.   Moreover, the $\hat \kappa_a^2\, \hat \kappa_b^2$ dependence is transparent in these plots. $\Delta B_d$ peaks at $(\kappa_1/\kappa_3,\kappa_2/\kappa_3)\sim (\pm 1,0)$, while  $\Delta B_s$ peaks at $\sim (0,\pm 1)$.  For $\Delta m_K$ and $\Delta m_D$, we can see that moving from $\kappa_1=\kappa_2 =\kappa_3$ to the corners, constraints rise by the expected factor of $\sim 16/9$.\footnote{Note that $\Delta m_D$ is rotated by the Cabbibo angle $\theta_c$ relative to the other observables. This is a consequence of the   the fact that, as discussed in  in Appendix \ref{app:scan}, our $\kappa_1,\, \kappa_2,$ and $\kappa_3$ directions are chosen to align with the low energy down, strange and bottom quark. For $\Delta m_D$, the dependence on $\kappa_i$ proceeds through the LL sector of the up-squark mass matrix, where there is an additional rotation by $V_{CKM}$.}
    
These results from meson mixing may seem to conflict with the SUSY flavor problem, which suggests that, with $\order{1}$ flavor-violation, the SUSY scale needs to enter above $\sim 500$~TeV due to constraints from $\Delta m_K$ and $\Delta m_D$  \cite{Altmannshofer:2009ne, Altmannshofer:2013lfa}.   However, these constraints are driven by the $C_S^{LR}$ Wilson coefficient, while \chifv\ only generates the $C^{LL}_V$ operator.  In the MSSM, the contribution of the latter to $\Delta m_{K,D}$ is suppressed by $\sim 10^{-4}$\,--\,$10^{-3}$ relative to the former. This is due to three separate effects that all work in the same direction.  First, the hadronic matrix elements differ between these two operators,  
\be
\frac{\left< K | S_{LR} | \bar K\right>}{\left< K | V_{LL} | \bar K\right>} =\frac 34 \frac{B_S^{LR}}{B_V^{LL}} R_K  \sim 35,\qq \frac{\left< D | S_{LR} | \bar D\right>}{\left< D | V_{LL} | \bar D \right>} =\frac 34 \frac{B_S^{LR}}{B_V^{LL}} R_D  \sim 4.
\ee   
Next, the SUSY contributions to the Wilson coefficients are also quite different in size.   From the MIA with $\order{1}$ mass-insertions, it is easy to see that $C_S^{LR}/C^{LL}_V\sim 30$ \cite{Altmannshofer:2009ne}.  Lastly, the QCD running from the SUSY scale to $\sim2$ GeV suppresses $C^{LL}_V$ by 20-30\%, but \emph{enhances} $C_S^{LR}$ by a factor of $\sim 3$ \cite{Buras:2001ra}.  All of these factors conspire to drop the scale of sensitivity to $\Delta m_K$ and $\Delta m_D$ to the TeV scale in  \chifv\ models.

\subsubsection{$K^\pm\rightarrow \pi^\pm \nu \bar\nu$}

 Unlike meson mixing in the previous subsection, the $K^\pm \rightarrow \pi^\pm \nu \bar\nu$ observable  (along with all other $\Delta F=1$ observables) enters as a matrix element squared, and thus interference with the standard model contribution can be important.  The expression for this branching ratio is (see appendix~\ref{app:Kpvv}),
\beq
\label{KtopinunuBR}
\text{BR}(K^\pm\rightarrow \pi^\pm \nu \bar\nu)=  c_+ v^4 \left| {C}^{LL}_{V,SM}+{C}^{LL}_V + {C}^{RL}_V\right|^2
\eeq
where $ c_+ v^4 = 4.9\times 10^{9} \gev^4$, $C^{LL}_{V,SM} = (-1.21+0.39 i)\times 10^{-10} \gev^{-2}$, and all EGMSB effects are contained in ${C}^{LL}_V$ and ${C}^{RL}_V$. These are the Wilson coefficients from the $\Delta F=1$ effective Hamiltonian built out of the four-fermi operators
\bea
\label{fourfermiDF1}
& (\CO_S^{MN})_{ab} = ( \bar q_a   P_M q_b) ( \bar\ell   P_N \ell)\\
& (\CO_V^{MN})_{ab} = ( \bar q_a \gamma^\mu P_M q_b) ( \bar \ell \gamma_\mu P_N \ell)\\
& (\CO_T^{MN})_{ab} = ( \bar q_a \sigma^{\mu \nu}P_M q_b) ( \bar \ell \sigma_{\mu \nu} P_N \ell)
\eea
Here $ab=12$ for $K^\pm\to\pi^\pm\nu\bar\nu$. In the MSSM, these Wilson coefficients arise through one-loop box and $Z$-penguin diagrams.

\begin{figure}[!t]
\begin{center}
\includegraphics[width=7.5cm,angle=0]{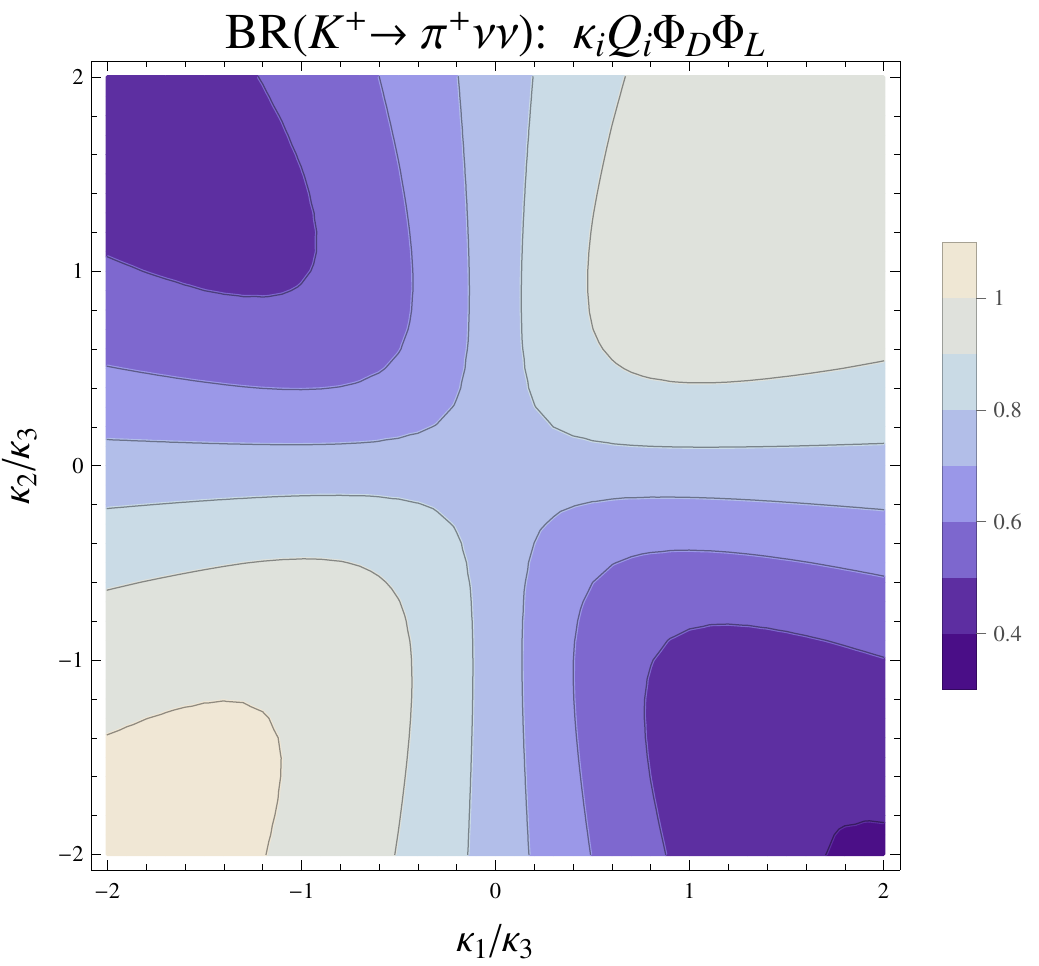}
\end{center}
\caption{Contours of the difference in BR$(K^+\to\pi^+\nu\bar\nu)$ between experiment and the SM + EGMSB predictions, given in units of the net uncertainty (added in quadrature).  A contour of 2 corresponds roughly to a 95\% exclusion.}
\label{fig:kpvv}
\end{figure}

The full  \FormFlavor\ result is shown in figure~\ref{fig:kpvv}. The general trend of $K^\pm \rightarrow \pi^\pm \nu \bar\nu$ can again be understood through use of the features of rank 1 \chifv\ discussed in section \ref{rank1miasec}. First of all, as for the $\Delta F=2$ observables, $C^{RL}_{V,ab}$ must be zero in the third-generation dominant approximation -- since it transforms in the adjoint+singlet of $SU(3)_D$, using the available spurions one can only obtain something that is nonzero in the 33 component. Thus, we can focus on $C^{LL}_{V,ab}$. This transforms in the adjoint+singlet of $SU(3)_Q$. Using flavor violation in the LL block only, the form of $C^{LL}_{V,ab}$ is constrained by the symmetries to be:
\begin{equation}
\label{CVLMdeltaF1}
C^{LL}_{V,ab} = {\hat\kappa_a \hat\kappa_b\over m_{S}^2} f_2  +\dots 
\end{equation}
where $\dots$  contains higher orders in $v/m_{SUSY}$ and other irrelevant terms, and $f_2$ is a dimensionless function  of superpartner mass ratios. We note that (\ref{CVLMdeltaF1}) can come from box diagrams or $Z$-penguin diagrams, but in the latter case, the $1/m_Z^2$ from the $Z$ propagator must be canceled out by two insertions of wino-Higgsino mixing. Insertions of LR mixing from the squark mass matrix would also cancel out the $1/m_Z^2$, but the heavy right-handed squark masses suppress these contributions enough that they may be ignored.

Now we will compare against explicit computations of the Wilson coefficients. We confirm that the $C_V^{RL}$ coefficients are all negligible. For $C_V^{LL}$, we find that the chargino diagrams dominate, and they are given by 
\beq
C^{LL}_V = \hat\kappa_1\hat\kappa_2 \frac{\alpha_2^2}{12 m_{S}^2}f_{\tilde{\chi}^\pm}^{K\rightarrow \pi \nu \nu}(x_l,x_2,x_\mu)
\eeq
Again, $f_{\tilde{\chi}^\pm}^{K\rightarrow \pi \nu \nu}(x_\ell,x_2,x_\mu) \sim 1.5$ is a loop function of the sparticle mass-ratios $x_\ell=m_{\tilde{\ell}}^2/m_{S}^2$, $x_2=M_2^2/m_{S}^2$, and $x_\mu=\mu^2/m_{S}^2$, defined in appendix~\ref{app:Loops}, and varies only mildly across the parameter space.    This is fully consistent with the general form (\ref{CVLMdeltaF1}). This also explains why the charginos dominate over the gluinos, since the gluinos can only give rise to $Z$-penguin diagrams that are nonzero by virtue of down sector LR squark-mixing insertions.

Substituting in numerically for the loop function, $m_S$ and $\alpha_2$, we find 
\beq
C^{LL}_{V}\sim \lp 8 \times 10^{-11} \mbox{ GeV}^{-2}\rp \hat\kappa_1\hat\kappa_2 
\eeq
This simple function of $\vec\kappa$, when added to the SM contribution and substituted into (\ref{KtopinunuBR}), reproduces well the features of fig.~\ref{fig:kpvv}. It grows in magnitude fastest along the lines $\kappa_1=\pm\,\kappa_2$, asymptoting to the values $\pm\, 4\times 10^{-11}$~GeV$^{-2}$. In the corners of the parameter space we obtain a deviation from the SM prediction of
\beq
\Delta\text{BR}(K^+\rightarrow \pi^+ \nu \bar \nu) =  c_+ v^4 \lp2\Re\left[C^{LL}_{V,SM} C^{LL*}_V \right]+\abs{C^{LL}_V}^2\rp \approx  \mp\, 5\times 10^{-11},
\eeq
This is smaller than the current experimental+theoretical uncertainty shown in table \ref{tab:obs}. However, because the SM prediction for the BR is a little lower than the experimentally observed value, moving along the $\kappa_1=+\kappa_2$ line slightly exacerbates the difference, while moving along the $\kappa_1=-\kappa_2$ line slightly lessens it.

\subsubsection{$b \rightarrow \,s \gamma$ and $b \rightarrow \,d \gamma$}

\begin{figure}[!t]
\begin{center}
\includegraphics[width=7.5cm,angle=0]{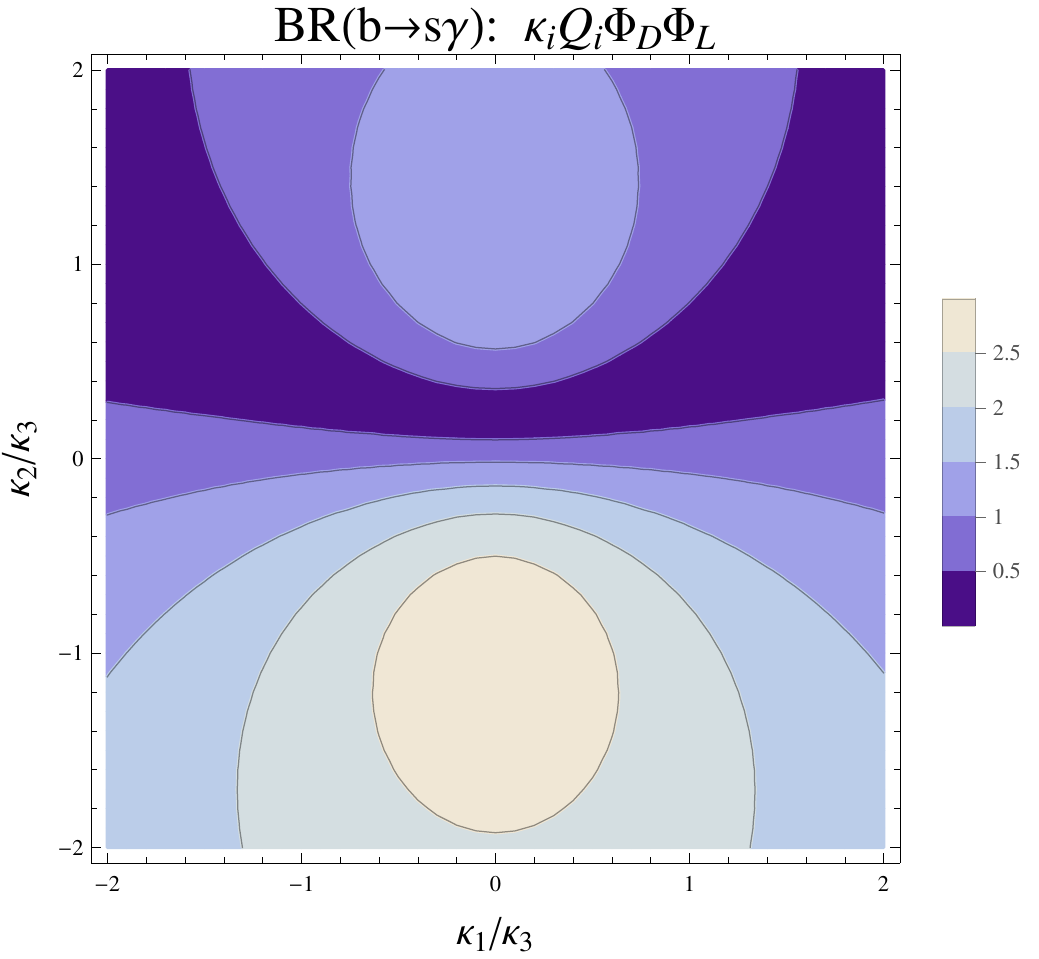}
\includegraphics[width=7.5cm,angle=0]{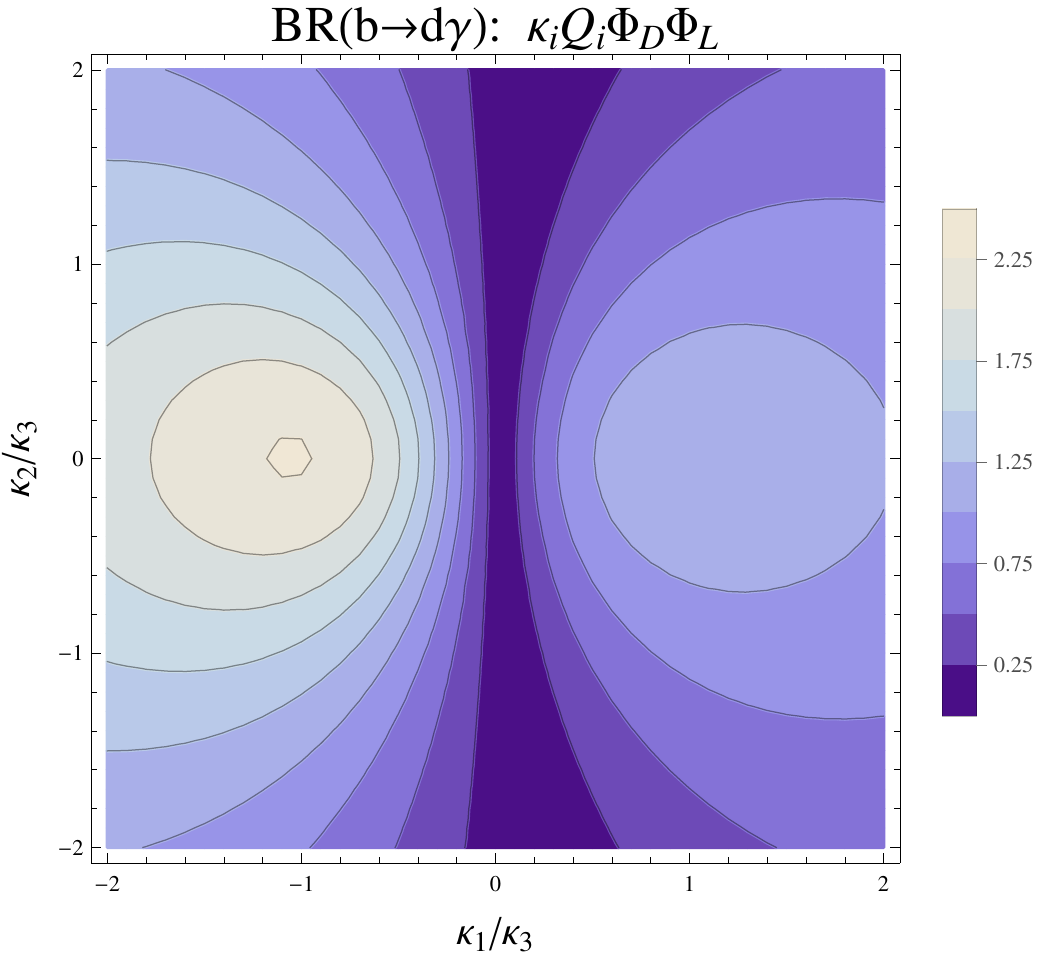}
\end{center}
\caption{Sensitivity to $b\to s\gamma$ (left) and $b\to d\gamma$ (right), same conventions as fig.~\ref{fig:kpvv}. }
\label{fig:bsg}
\end{figure}

 The expression for the $b \rightarrow \,s \gamma$ branching ratio (see appendix \ref{app:bsg}) is,
\be\label{btosgammaBR}
\text{BR}(b\rightarrow s \gamma)=  c_\gamma v^2 \Big(\lvert C^L_{A,SM} + C^L_A \rvert^2 + \lvert C^R_A \rvert^2 \Big),
\ee
where $ c_\gamma v^2=1.97\times 10^{12}\gev^2$ and $C^L_{A,SM} \sim 1.3 \times 10^{-8} \gev^{-1}$. The new physics is contained in $C_A^L$ and $C_A^R$; these come from the effective Hamiltonian built out of the  dimension 5 operators
\begin{equation}
 (\CO_A^{M})_{ab} = e ( \bar q_a \sigma^{\mu \nu}P_M q_b) F_{\mu\nu}
\end{equation}
where $ab=32$ for $b\to s\gamma$. These operators require $SU(2)$ breaking, so their Wilson coefficients are zero in the $v\to 0$ limit. Naively, this would mean that the result is negligible in our zeroth-order expansion in $v/m_{SUSY}$. However, the SM Wilson coefficient is suppressed by $m_b/v$, so here we consider one higher order in the $v/m_{SUSY}$ expansion in order to capture a numerically relevant result.

Since $C_A^L$ ($C_A^R$) transforms in the $({\bf 3},\bar {\bf 3})$ of $SU(3)_D\times SU(3)_Q$ ($SU(3)_Q\times SU(3)_D$), it must be proportional to $m_d$ acting on the left (right).   The latter is zero in the third-generation dominant approximation, so it suffices to focus on $C_A^L$. With just left-handed squarks propagating in the loop, the Wilson coefficient must be given by:
\beq
\label{CLAab}
C^{L}_{A,32} ={(m_d K  \tan\beta)_{32}\over \kappa^2} {1\over m_{S}^2}  f_6 =  \hat\kappa_2\hat\kappa_3{m_b \tan\beta\over m_S^2} f_6
\eeq
Here we assumed that $m_d$ is accompanied by a $\tan\beta$ enhancement, otherwise the entire effect is numerically negligible.

Comparing with explicit calculation, we find again that charginos give the dominant contribution to the Wilson coefficient, through the quark-squark-Higgsino vertex. (Contributions of the form (\ref{CLAab}) can also arise through gluino and neutralino loops, but here the factor of $m_b\tan\beta$ arises through the LR block of the down-squark mass matrix, so the diagrams are suppressed by heavy right-handed squarks propagating in the loop.)  The dominant chargino diagram gives, 
\beq
C^L_A=   \hat\kappa_2\hat\kappa_3 {m_b \tan\beta \over m_{S}^2} \frac{11\alpha_2  }{288 \pi  }f_{\tilde{\chi}^\pm}^{ b \rightarrow s/d\,\gamma,\text{\tiny{peng}}}(x_\mu,x_2),
\label{eq:bsgCLA}
\eeq
where $x_\mu= \mu^2/m_{S}^2$, $x_2=M_2^2/m_{S}^2$, and $f_{\tilde{\chi}^\pm}^{ b \rightarrow s/d\,\gamma,\text{\tiny{peng}}}(x_\mu,x_2)\sim 0.5$ is a loop function defined in appendix~\ref{app:Loops}. This is fully in agreement with the general result (\ref{CLAab}). QCD running from the SUSY scale to the $b$ pole \cite{Grinstein:1990tj}  induces a mild 20\% suppression to (\ref{eq:bsgCLA}), and once again, we find this simple result is enough to account for the features of the  \FormFlavor\ plot in figure~\ref{fig:bsg}.

 Substituting in numerically for our parameter space (we take $\tan\beta=10$ as in \cite{Evans:2013kxa}), we obtain:
\beq
C_A^L \sim \lp3\times 10^{-9} \mbox{ GeV}^{-1}\rp  \hat\kappa_2\hat\kappa_3
\label{eq:bsgCLAsimp}
\eeq
Clearly its effects are largest when $\kappa_1=0$, and along the $\kappa_2$ axis it has a maximum (minimum) at $\kappa_2/\kappa_3=1$ (-1), while asymptoting back to zero as $\kappa_2/\kappa_3\to\pm\infty$. 
Substituting this into (\ref{btosgammaBR}) together with the SM Wilson coefficient, we find at these extrema a deviation from the SM prediction of
\beq
\left[\Delta\text{BR}(b\to s\gamma)\right]_{\kappa_2 =\pm\kappa_3} \sim   c_\gamma v^2 \lp2\Re\left[ C^L_{A,SM} C^{L*}_A \right]\rp \sim \pm\, 7\times 10^{-5},
\eeq
 Since the uncertainty on the measurement (combining the theoretical and experimental in quadrature) is $3.5\times 10^{-5}$, regions of exclusion are to be expected for $b\to s \gamma$. 

In practice, \FormFlavor\ does find a region of exclusion in the bottom half of the parameter space, shown in figure~\ref{fig:bsg}.  This exclusion is in part due to the current $\order{1\sigma}$ excess in the experimental measurement relative to the theoretical prediction,\footnote{Recently, an improved theoretical prediction of $b\to s \gamma$ was released \cite{Misiak:2015xwa}.  This work predicts that BR$(b\to s \gamma)=(3.36\pm0.23)\times 10^{-4}$, which is significantly more in line with the experimental value.  Unfortunately, insufficient details are provided in that work for us to modify \FormFlavor\ to account for this prediction.} which allows the $\sim 2\sigma$ change to constrain a large region of negative $\kappa_2$ in the plot.  Uncertainty on this observable is comparable in size between theory and experiment, so improvements on either side could make this observable more constraining and in a direction that other observables currently have no sensitivity.  

The observable $b\to d\gamma$ could also potentially place constraints.  The EGMSB contribution to the Wilson coefficient is the same as in (\ref{eq:bsgCLA})--(\ref{eq:bsgCLAsimp}), but with $\kappa_2 \leftrightarrow \kappa_1$. However, the SM contribution is of a different size, $C^L_{A,SM} \sim -\lp 2.4+1.1i\rp \times 10^{-9} \gev^{-1}$.  Importantly,  this is not much larger than the new physics contribution, so interference is very important.   This yields an approximate deviation of,
\bea
\left[\Delta\text{BR}(b\to d\gamma)\right]_{\kappa_1 =\kappa_3} \sim&\, c_\gamma v^2 \lp-2\abs{\Re\left[ C^L_{A,SM} C^{L*}_A \right]}+\abs{C^{L}_A}^2\rp \sim -0.8 \times 10^{-5} \\
\left[\Delta\text{BR}(b\to d\gamma)\right]_{\kappa_1 =-\kappa_3} \sim &\, c_\gamma v^2 \lp2\abs{\Re\left[ C^L_{A,SM} C^{L*}_A \right]}+\abs{C^{L}_A}^2\rp \sim 1.6 \times 10^{-5},
\eea
while the net uncertainty is $6.3\times 10^{-6}$.  Again, a two sigma exclusion could be possible, but it would only be expected near $\frac{\kappa_1}{\kappa_3}=-1$ due to the interference acting constructively.  This is borne out by figure~\ref{fig:bsg}, which shows a maximum deviation of $\approx 2.3$ times the net uncertainty.  Because the measurement of $b\to d \gamma$ is fairly recent, a future improvement that pushes the experimental uncertainty to the 10-15\% level would allow for $b\to d\gamma$ to place much tighter constraints on the region of  constructive interference, i.e., $\kappa_1<0$.

\subsubsection{$B_q \rightarrow \, \mu^+ \mu^-$}

\begin{figure}[!t]
\begin{center}
\includegraphics[width=7.5cm,angle=0]{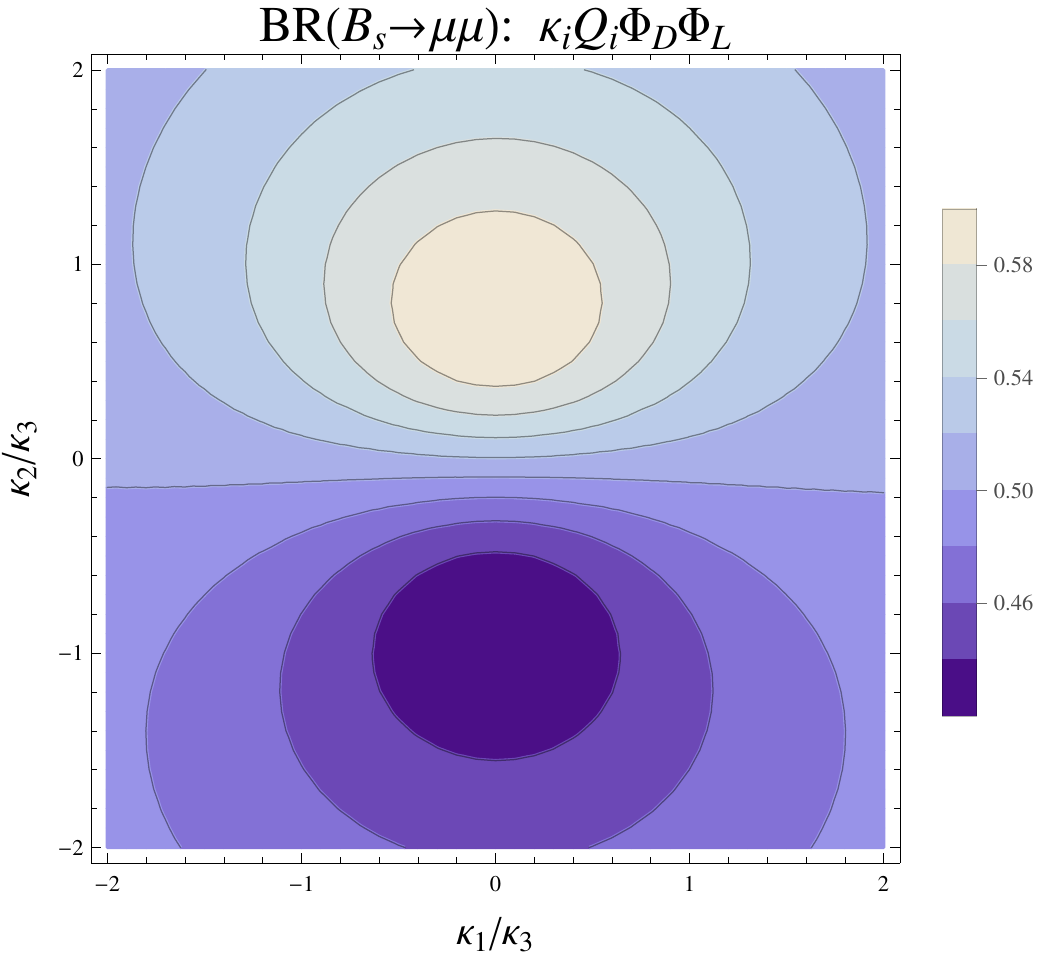}
\includegraphics[width=7.5cm,angle=0]{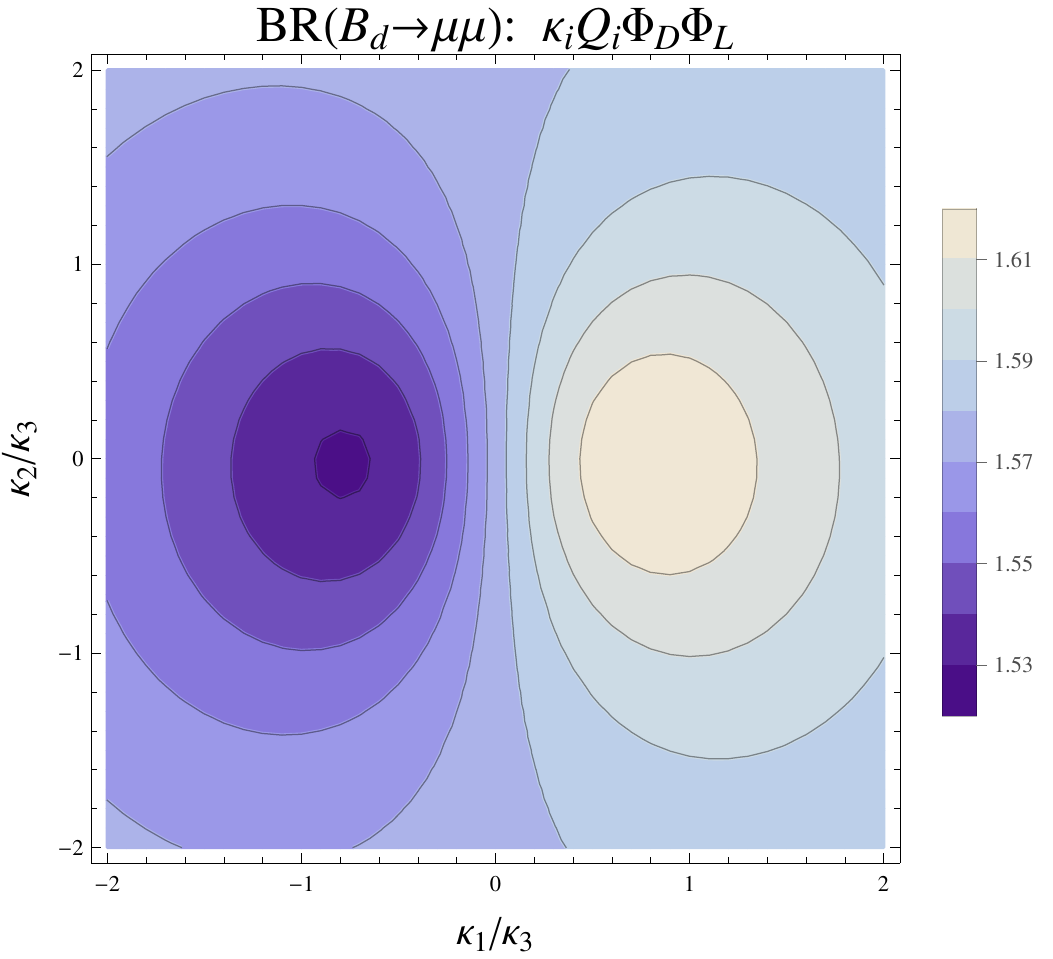}
\end{center}
\caption{Sensitivity to $B_s\to \mu\mu$ and $B_d\to \mu\mu$, same conventions as fig.~\ref{fig:kpvv}.  Neither is constraining over the parameter space.}
\label{fig:bqtomumu}
\end{figure}

As shown in fig.~\ref{fig:bqtomumu}, neither $B_s \rightarrow \, \mu^+ \mu^-$ nor $B_d \rightarrow \, \mu^+ \mu^-$ is at all close to constraining our EGMSB models. We include a brief discussion of these observables just for completeness sake. 

The branching ratio for $B_{q}\to\mu^+\mu^-$ is \cite{Bobeth:2002ch,Dedes:2008iw} (see appendix \ref{app:bmm} for more detailed formulas): 
\begin{equation}
\text{BR}(B_q\rightarrow \mu^+ \mu^-) = X_q \left\{  \left(1-\frac{4 m_\mu^2}{m_{B_q}^2}\right) | {F}^{(q)}_S |^2 + | {F}^{(q)}_P + {F}^{(q)}_A |^2\right\}
\end{equation}
where $X_s=5.36\times 10^7$ and $X_d=3.97\times 10^7$,
and
\bea
\label{eq:bmmFdef}
{F}^{(q)}_S =&\;  {m_{B_q}^3\over m_b+m_q} ({ C}_S^{ LL}+{ C}_S^{ LR}-{ C}_S^{ RR}-{ C}_S^{ RL}), \\ 
{F}^{(q)}_P =&\;  {m_{B_q}^3\over m_b+m_q}  (-{ C}_S^{ LL}+{ C}_S^{ LR}-{ C}_S^{ RR}+{ C}_S^{ RL}), \\ 
{F}^{(q)}_A =&\; 2m_{B_q}m_\mu ({ C}_V^{ LL}-{ C}_V^{ LR}+{ C}_V^{ RR}-{ C}_V^{ RL}),
\eea
Here the $C_Y^{MN}$ coefficients are those for the $\Delta F=1$ effective Hamiltonian introduced in  (\ref{fourfermiDF1}).  The three-loop Standard Model contribution \cite{Hermann:2013kca} is $F^{(d)}_{A,SM}=(1.5-0.6 i)\times 10^{-9}$ and $F^{(s)}_{A,SM}=(-7.9-0.1 i)\times 10^{-9}$.

\begin{figure}[!t]
\begin{center}
\includegraphics[width=10cm,angle=0]{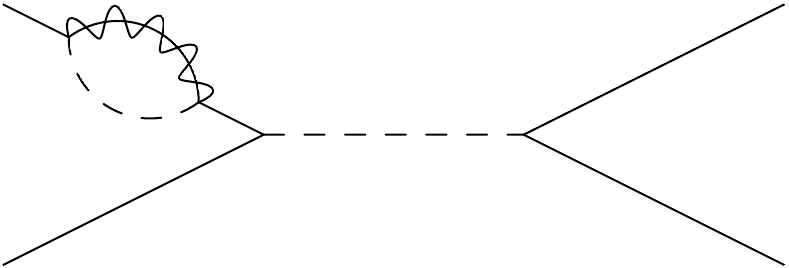}
\begin{picture}(0,0)(0,0)
\put(3,-2){$\mu^+$}
\put(3,94){$\mu^-$}
\put(-159,55){$A_0,H_0$}
\put(-300,94){$\bar q$}
\put(-300,-2){$b$}
\end{picture}
\end{center}
\caption{Wave function correction diagrams that provide the leading, $\tan^3\beta$ contributions to $B_q\to\mu^+\mu^-$.  Loop can be a gluino, chargino, or neutralino.}
\label{fig:Zcorrbsmumu}
\end{figure}
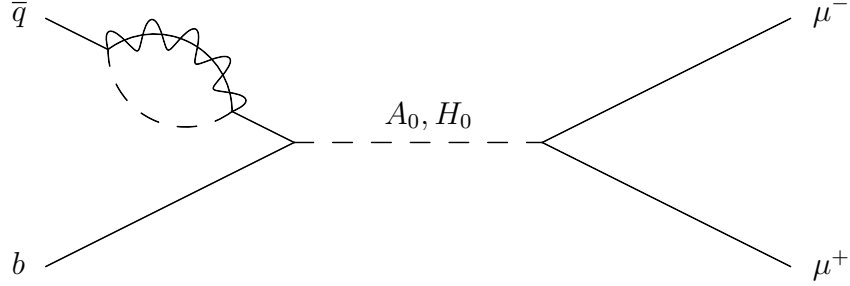

In the MSSM, the dominant contributions to $B_q\to\mu^+\mu^-$ are the $\tan^3\beta$ enhanced wave function correction diagrams with a heavy higgs propagator, see figure \ref{fig:Zcorrbsmumu}.  Here gluinos, charginos and neutralinos can run in the loop, with both CP even and odd higgs states along the penguin line.  For simplicity, we will quote the result only for gluinos; the answer for the others is very similar. As our higgs states are heavy, we use the relation $m_A^2\approx m_H^2$ to simplify expressions.  

These diagrams contribute to $C_S^{MN}$, and the discussion is similar to that of $b\to q\gamma$. $C_S^{LN}$ transforms in the $({\bf 3},{\bf \bar 3})$ of $SU(3)_D\times SU(3)_Q$ while $C_S^{RN}$ transforms in the $({\bf 3},{\bf \bar 3})$ of $SU(3)_Q\times SU(3)_D$. Thus as before, the latter is zero thanks to \chifv\ and we can focus on the former. Here we can afford to work at $v=0$ since we are not concerned with $Z$-penguins. With only left-handed squarks propagating in the loops, by symmetries the answer must be of the form:
\beq
\label{CLMSab}
C^{LM}_{S,3a} =  \frac{(K y_d)_{3a}  }{\kappa^2 m_{S}^2} f_5=   \hat\kappa_3\hat\kappa_a {y_b\over m_S^2} f_5
\eeq
where $a=2$ for $B_s\to\mu\mu$ and $a=1$ for $B_d\to\mu\mu$. Now the Yukawa coupling needed on symmetry grounds arises from the Higgs-quark-quark coupling. Explicit computation gives
\bea
\label{CSLRhch}
C_{S,\go}^{LR} &=\hat\kappa_3\hat\kappa_a  {m_b\tan\beta\over m_W} {m_\mu \tan^2\beta\over m_W} \frac{4\alpha_2 \alpha_s }{3 m_A^2 } \frac{\mu}{m_S}  f_{\tilde{g}}^{ B_q \rightarrow \mu^+ \mu^-,\text{\tiny{h-peng}}}(x_q,x_g)
\eea
where $x_g= m_{\tilde{g}}^2/m_{0}^2$ and $x _{q}=m_{S}^2/m_{0}^2$ and the loop function (see appendix~\ref{app:Loops}) is $f_{\tilde{g}}^{ B_q \rightarrow \mu^+ \mu^-,\text{\tiny{h-peng}}}(x_q,x_g)\sim 0.1$. Since $m_A\sim m_{SUSY}$, this leading contribution is of the form (\ref{CLMSab}) as expected. 

From the above expressions, we can translate to the phenomenologically useful parameters
\bea
& {F}^{(s)}_P = \frac{m_{B_s}^3}{m_b+m_s} C_S^{LR} \sim  (2\times 10^{-10} )\hat\kappa_3\hat\kappa_2 \\
& {F}^{(d)}_P = \frac{m_{B_d}^3}{m_b+m_d} C_S^{LR} \sim  (2\times 10^{-10} )\hat\kappa_3\hat\kappa_1 .
\label{eq:bsmumuFs}
\eea
 These translate into maximum deviations from the SM branching fractions by
\bea
\left[\Delta\text{BR}(B_s\rightarrow \mu^+ \mu^-)\right]_{\text{\tiny{MAX}}}&\simeq X_s \left\{ 2 \Re  \left[ F^{(s)*}_{A,SM}  {F}^{(s)}_P \right]\right\} \sim 9\times 10^{-11}\\
\left[\Delta\text{BR}(B_d\rightarrow \mu^+ \mu^-)\right]_{\text{\tiny{MAX}}}&\simeq X_d \left\{ 2 \Re  \left[ F^{(d)*}_{A,SM} {F}^{(d)}_A\right]\right\} \sim 1.2 \times 10^{-11}
\eea
 These deviations are an order of magnitude smaller than the uncertainties on their respective measurements, and therefore neither $B_s\rightarrow \mu^+\mu^-$ nor $B_d\rightarrow \mu^+\mu^-$ place any meaningful constraints on our models.\footnote{In order for $B_s\to\mu^+\mu^-$ to place any meaningful constraint, one would have to take $\tan\beta$ much larger than the $\tan\beta\sim 10$ that we have assumed in this paper, e.g., $\tan\beta\gtrsim 30$. }

\subsection{U-class Models}

As discussed in section \ref{sec:chiFV}, all $U$-class \chifv\ models receive very few constraints from flavor observables.   First, according to the \chifv\ texture of $U$-class models (\ref{eq:Utexturegenu}), all flavor-violation is restricted to the up-squark sector only.  As most potentially constraining flavor observables have external down-type quarks,  chargino diagrams are required for sensitivity to flavor-violation.  However, as is further shown in (\ref{eq:Utexturegenu}), there is no flavor-violation in the ${LL}$ block; thus pure wino diagrams cannot contribute.  Higgsino diagrams introduce Yukawa couplings which suppress contributions to down sector observables enough that none of these would be remotely constraining in the foreseeable future. We have verified all of these general results in the context of our type I EGMSB models.

\begin{figure}[!t]
\begin{center}
\includegraphics[width=7.5cm,angle=0]{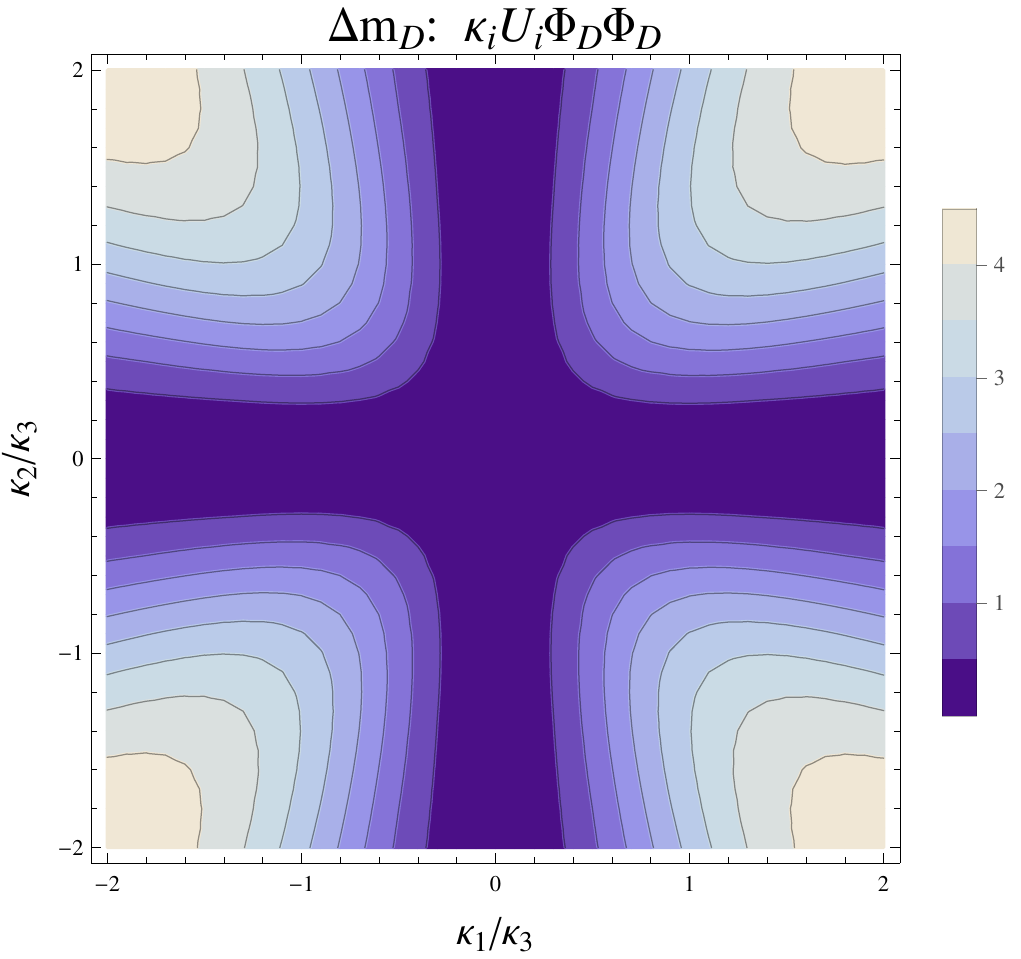}
\caption{Plot of $\Delta m_D$ meson mixing observable for a $U$-class model.}
\label{fig:DmxU}
\end{center}
\end{figure}

There can be constraints from the $D$-meson system.  Contributions to $\Delta m_D$ are as in the $Q$-class models, only now with the $C^{RR}_V$ contribution  dominating, so we expect very similar constraints from $\Delta m_D$. This is indeed shown in figure \ref{fig:DmxU}.  Again, although the contour of 2 does not represent a 95\% confidence level exclusion, values larger than 1 necessitate a cancellation between the standard model and the new physics contributions.  It should be noted that a viable possibility is that the standard model contribution is in fact much smaller than the observed value, and the contour of one is actually where EGMSB entirely accounts for the $\Delta m_D$ measurement.  As with the $Q$-class models, the sensitivity vanishes as either $\kappa_1\to0$ or $\kappa_2\to0$, and increases most rapidly along the diagonals.

\section{Conclusions}
\label{sec:conclusions}

\subsection{Summary and Discussion}

In this work, we performed a detailed investigation into the precision flavor constraints on extended GMSB models.  These models, where the usual GMSB contributions are augmented by direct matter-messenger couplings, are well motivated in light of the recent discovery of a Higgs boson near $125\,\rm{GeV}$. However, since these models are not necessarily MFV,  in their full, three-family generalizations they could be potentially dangerous from the point of view of flavor. 

Our work required a computer program that could turn general MSSM spectra into precision flavor observables. We found that existing programs had various limitations -- they either assumed MFV, were numerically unstable, or had incorrectly transcribed formulas from the literature.  This motivated us to develop \FormFlavor, a comprehensive package that computes flavor observables \textit{ab initio} starting from the Feynman rules, and uses a modular framework that enables us to add new observables in a uniform and straightforward way. 

Using \FormFlavor, we studied the flavor constraints on the three-family generalizations of the EGMSB models  of \cite{Evans:2013kxa}. The results we encountered from this systematic study were interesting and unexpected. We found that despite the introduction of $\order{1}$ flavor-violating couplings, there are currently very few constraints on EGMSB models. To validate the numerical results of \FormFlavor, we compared them in detail with analytic formulas for the Wilson coefficients derived using a combination of flavor symmetry arguments and direct calculation, and we found excellent agreement. 

\begin{figure}[lt]
\includegraphics[width=6.9cm,angle=0]{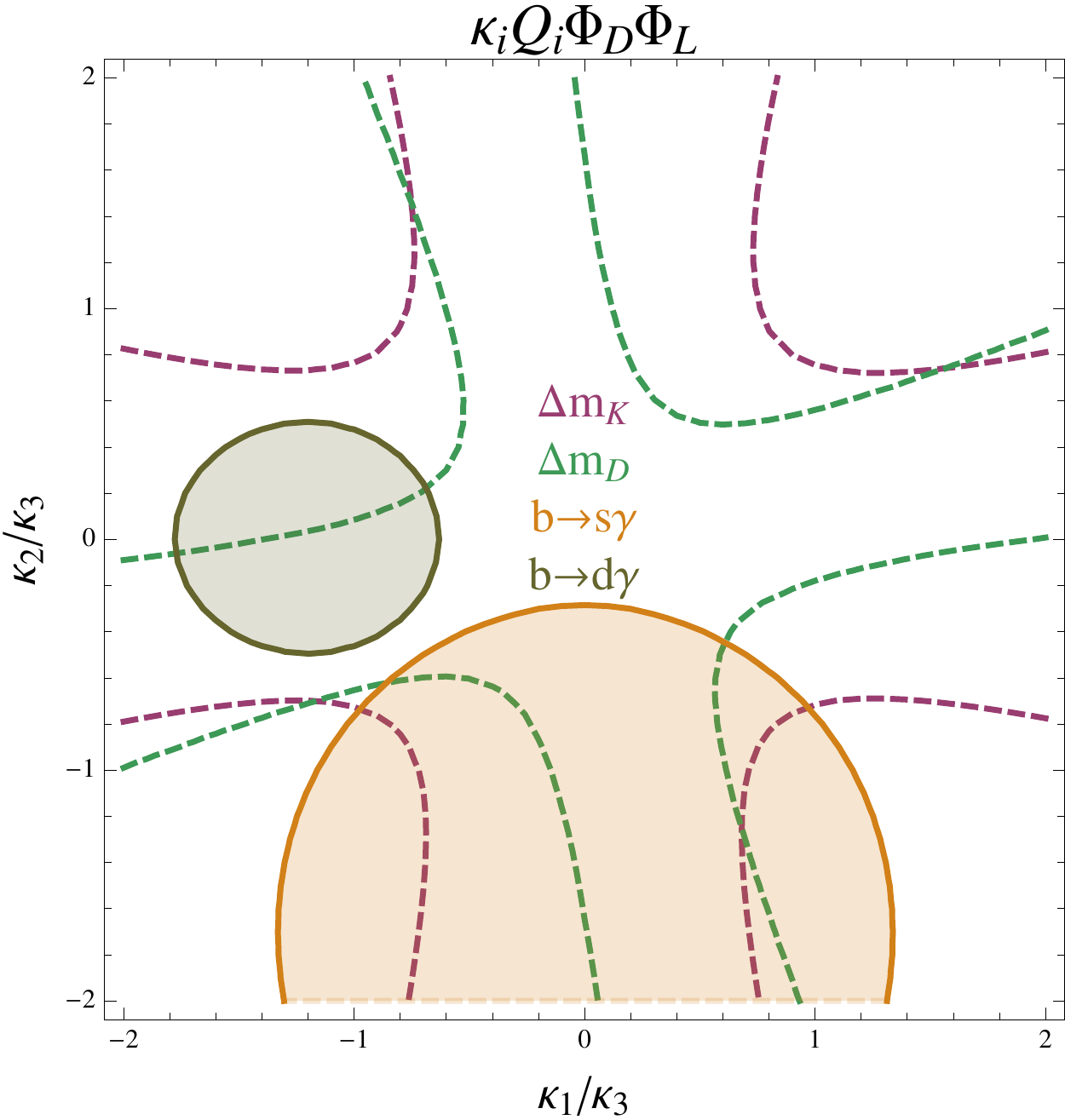} \qqq
\includegraphics[width=6.9cm,angle=0]{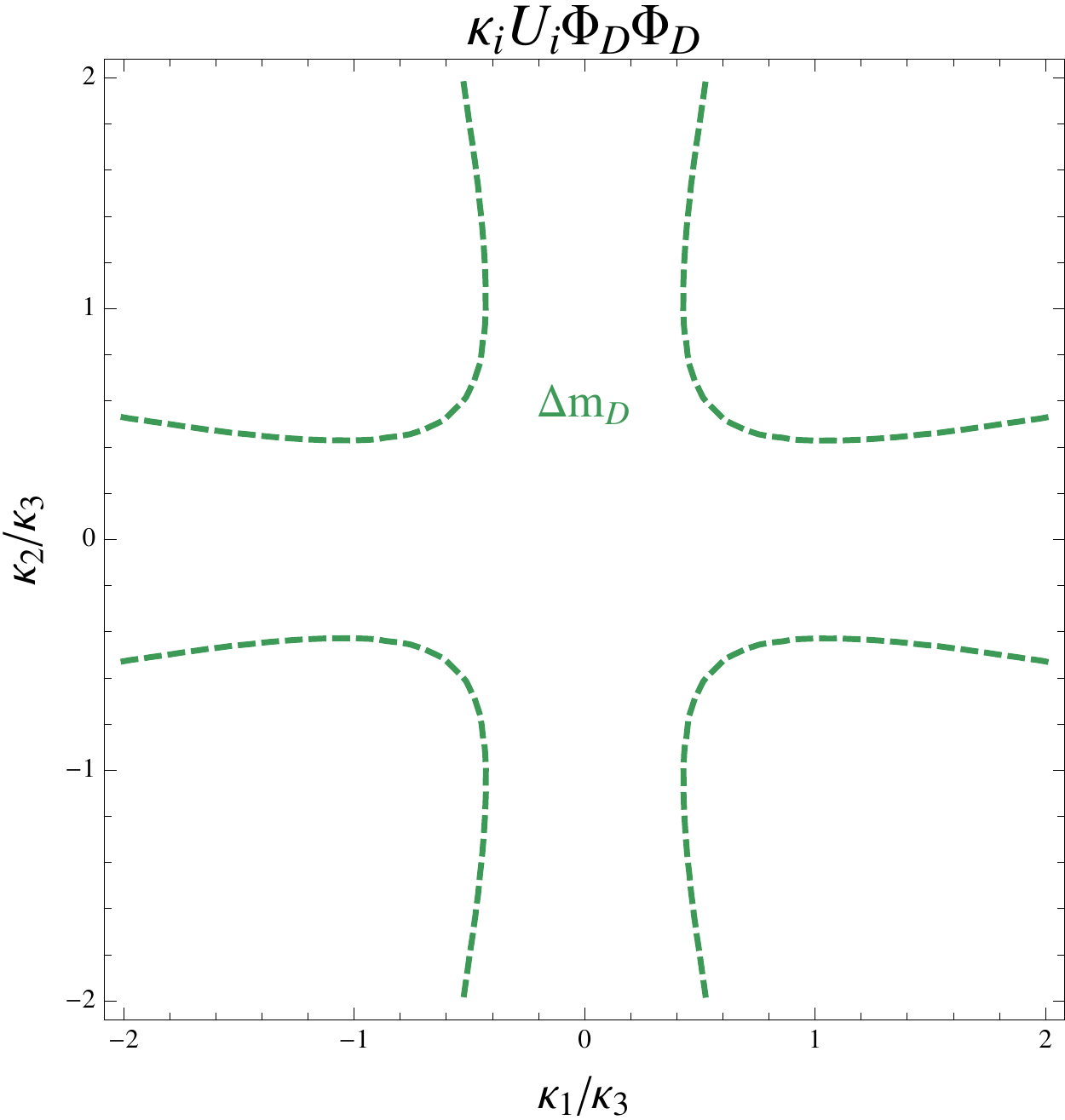}
\caption{Flavor sensitivity for a $Q$-class (left) and $U$-class (right) model.   Shaded regions of this EGMSB parameter space are excluded to 95\% confidence level by current measurements of flavor observables as described in the text.  Dashed lines indicate the locations where the EGMSB contribution to $\Delta m_K$ and $\Delta m_D$ are equal to the measured value (as these currently cannot provide a genuine exclusion due to poorly controlled theoretical uncertainties).   In both cases, observables not shown are not constraining.}
\label{fig:constraints}
\end{figure}

The mild flavor constraints in these models are illustrated in the summary plot of figure~\ref{fig:constraints}.   $U$-class models only receive constraints from $\Delta m_D$.  $Q$-class models are constrained by $\Delta m_K$ and $\Delta m_D$ in the corners of the plot, while the radiative $b\to s\gamma$ and $b\to d\gamma$ each exclude a single bubble near $\kappa_2 = -\kappa_3$ and $\kappa_1 = -\kappa_3$, respectively.  We note that the excluded region from $b\to s\gamma$ is largely due to the current $\sim1\sigma$ discrepancy between the theoretical prediction and the measurement, which new theoretical work suggests will disappear \cite{Misiak:2015xwa}.

The results from these models may seem at odds with the SUSY flavor problem. We have argued that the mildness of the flavor violation in these models originates from the fact that they obey the ``chiral flavor violation" ansatz, whereby flavor is violated only by the Yukawas and spurions of a single $SU(3)$ of the full $SU(3)^5$ SM flavor symmetry.  We showed that \chifv\  prevents many of the most problematic contributions to flavor observables from arising in the MSSM, and allows for $\order{1}$ flavor-violation in EGMSB models.  

\subsection{Future Constraints}
\label{sec:future}

Although constraints are currently very mild,  there is immense potential to further probe these models in the near future. 
\begin{itemize}

\item For $\Delta m_K$, short-distance predictions exist with moderate uncertainty, i.e., $(3.1\pm1.2)\times 10^{-15}$\,\rm{GeV}~\cite{Brod:2011ty}, but the long-distance contributions are currently unknown.  However, an accurate, full calculation on the lattice, including both long- and short-distance contributions, may be coming in the near future, as promising preliminary work on the subject shows~\cite{Christ:2012se,Bai:2014cva}. 

\item  Expected incremental improvements to the theoretical uncertainty of the standard model prediction could make $\Delta B_d$ into a constraining observable soon.   Estimates suggest that both the bag parameter, $B_{V,B_d}^{LL}$, and the relevant CKM elements could be calculated on the lattice to significantly improved levels by 2018 \cite{Butler:2013kdw}.

\item Belle II \cite{Abe:2010gxa} is expected to make significantly improved measurements to both $b\to s \gamma$ and $b\to d \gamma$ \cite{Aushev:2010bq}, which would allow for both of these observables to constrain more of the parameter space.

\item  NA62 \cite{NA62TDR} at CERN is projected to be able to measure $\text{BR}(K^\pm\rightarrow \pi^\pm \nu \bar\nu)$ to a precision of about 10\% \cite{Fantechi:2014hqa} of the SM value.  This is an improvement of more than an order of magnitude relative to the current measurement.  In a few years, $K^+ \to \pi^+ \nu\bar\nu$ will be one of the the most sensitive flavor observable to $Q$-class models. Perhaps more importantly, if NA62 were to measure a deviation from the standard model prediction, this model would provide a significant motivation to invest in an experiment like ORKA \cite{E.T.WorcesterfortheORKA:2013cya}, that would be able to hone in on the parameter space.

\end{itemize}

\begin{figure}[lt]
\includegraphics[width=6.9cm,angle=0]{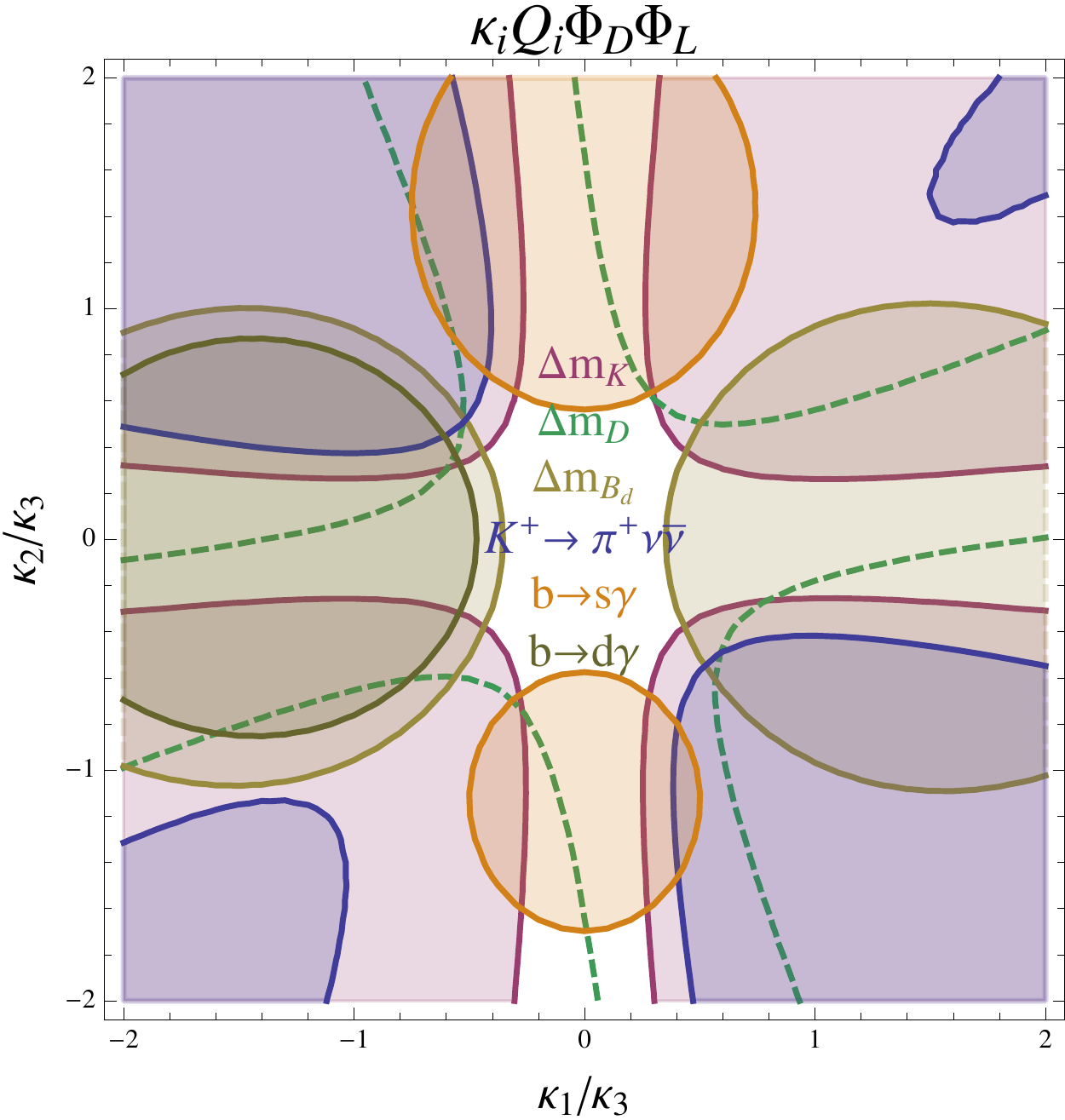}  \vspace{-69.3mm}

 \hspace{7cm} 
\begin{tabular}{|c|c|}
\hline
Observable &  Projected Accuracy \\
\hline 
$\Delta m_{K}$ & $10\%_{\text{\tiny{th.}}}$ \cite{Christ:2014qwa}  \\
\hline
$\Delta m_{B_d}$ & $10\%_{\text{\tiny{th.}}}$ \cite{Butler:2013kdw}  \\
\hline
$\Delta m_{B_s}$  & $5\%_{\text{\tiny{th.}}}$ \cite{Butler:2013kdw}\\
\hline
$\Delta m_{D}$ & None \\
\hline
 ${ Br}(K^+\to \pi^+ \nu \bar{\nu})$ & $10\%_{\text{\tiny{exp.}}}$ \cite{Butler:2013kdw}\\
\hline
${ Br}(B \to X_s \gamma) $ & $7\%_{\text{\tiny{exp.}}}$ \cite{Aushev:2010bq}\\ 
\hline
${ Br}(B \to X_d \gamma) $  & $24\%_{\text{\tiny{exp.}}}$ \cite{Aushev:2010bq} \\
\hline
${ Br}(B_s \to \mu^+ \mu^-)$  & $15\%_{\text{\tiny{exp.}}}$ \cite{Butler:2013kdw}\\ 
\hline
${ Br}(B_d \to \mu^+ \mu^-)$  & $35\%_{\text{\tiny{exp.}}}$ \cite{Butler:2013kdw}\\ 
\hline
\end{tabular} \vspace{3mm}

\caption{Future projections for flavor constraints on EGMSB in the $\sim$~3--5 year range.  In the figure, the errors have been updated to the fractional values in the accompanying table, and the central values have set equal between theory and experiment.  Depicted is the same $Q$-class model from figure~\ref{fig:constraints}.  The dashed green line, which is unchanged from figure~\ref{fig:constraints}, indicates where the EGMSB contribution to $\Delta m_D$ is equal to the measured value. Observables not shown are not constraining. }
\label{fig:future}
\end{figure}

Due to these potentially significant theoretical and experimental improvements, much of the parameter space in $Q$-class models could be probed in just a few years.  The projected sensitivities are shown in figure~\ref{fig:future}.   Although $Q$-class models have an exciting future in flavor, $U$-class models remain completely unconstrained.  Short of lattice predictions for $\Delta m_D$, no observables in the current program are sensitive to these these models.  Charm factories and precision top studies could someday explore this space in, for instance, $c\to u\gamma$ or $t\to c/u\gamma$.  However, for the moment, $U$-class models are resilient against flavor constraints.

\subsection{Future Directions}
\label{sec:futureus}

There are many avenues for future investigations. Here we list a few.

\begin{itemize}

\item As we alluded to earlier, the \chifv\ ansatz is a general paradigm, and it provides a novel and realistic solution to the SUSY flavor problem.  This texture, its possible origins, and consequences are worthy of further study \cite{ChiFVpaper}. 

\item Another interesting question is that of CP violation in these models.  The goal of this work was to focus on the SUSY flavor problem, and so the EGMSB couplings were intentionally assumed to be real. Allowing for $\arg\kappa_i\ne 0$ would give nontrivial contributions to CP-violating observables, some of which (such as $\epsilon_K$ and the neutron EDM) are typically extremely constraining.  An interesting question is to what extent the \chifv~texture protects EGMSB models from the SUSY CP problem before there is any conflict with data.  We plan to study this in detail in an upcoming work \cite{CPVpaper}.

\item The possibility of heavily mixed squarks allows for very interesting collider signatures~\cite{Mahbubani:2012qq,Blanke:2013uia,Agrawal:2013kha,Backovic:2015rwa}. These EGMSB models provide a flavor-safe proof-of-concept motivation for experimental searches at ATLAS and CMS.

\item While we made some effort to ensure that the flavor-violating $A$-terms do not destabilize the vacuum (see the discussion in appendix \ref{app:scan}), it 
would be interesting to study in more detail the vacuum stability of these flavor-violating EGMSB models, along the lines of \cite{Park:2010wf}.

\end{itemize}

\section*{Acknowledgments}
\noindent
We are especially grateful to W.~Altmannshofer for valuable discussions and comments on the draft. We also thank A.~El-Khadra, S.~Gori, S.~Martin, G.~Perez, K.~Pitts, and F.~Staub for useful discussions.  The work of J.A.E.\ was supported in part by the National Science Foundation under Grant No.~PHYS-1066293 and the hospitality of the Aspen Center for Physics. The work of D.S.\ was supported in part by a DOE Early Career Award and a Sloan Foundation Fellowship. The work of A.T.\ was supported in part by DOE grant DOE-SC0010008.

\appendix

\section{Details of the Deformation and Parameter Scan}
\label{app:scan}

In this appendix, we detail our parameter scan.  In particular, we specify the deformation about the best points in \cite{Evans:2013kxa} that we use.  Our type I models  are characterized by the following parameter space: 
\bea
\label{eq:parspace}
(\kappa_1,\kappa_2,\kappa_3,\Lambda/M,\Lambda)
\eea
In \cite{Evans:2013kxa}, the models were studied at the $\kappa_1=\kappa_2=0$ point. For each $\kappa_3$ and $\Lambda/M$, $\Lambda$ (which sets the overall scale of the superpartner spectrum) was increased until $m_h=125$~GeV was achieved. Then the fine-tuning of the point was estimated and the region of least fine-tuning in $(\kappa_3,\Lambda/M)$ space was identified. Our aim is to investigate these regions in the presence of nontrivial flavor violation, parametrized by $\kappa_1$ and $\kappa_2$. A full optimization of the fine-tuning would involve a five-dimensional scan in the $(\kappa_1,\kappa_2,\kappa_3,\Lambda/M, M)$ parameter space. Such an endeavor is not computationally feasible and moreover is unnecessary. The main question we intend to explore is how qualitatively dangerous the flavor-physics contributions are in these EGMSB models.  In principle, this question can be answered using a simpler scan where we essentially fix $(\kappa_3,\Lambda/M,M)$ as in \cite{Evans:2013kxa} and then vary $\kappa_1$, $\kappa_2$. However, there are a number of subtleties to take into account when doing so,  most of which stem from the fact that the points from \cite{Evans:2013kxa}\ tended to have light stops as a result of a cancellation (this is unsurprising due to the tuning involved). 

\begin{itemize}

\item  We must be careful to choose $(\kappa_3,\Lambda/M)$ such that the point at $\kappa_1=\kappa_2=0$ is unconstrained by Run I searches. In practice, to get a model that is not constrained by the LHC and has a good tuning value, we will lower $\frac \Lambda M$ away from the least-tuned point identified in \cite{Evans:2013kxa}.  This increases the fine-tuning required in the models only by about 10\%. The parameters we choose (at the origin) are $\kappa_3=0.858$, $N=6$, $\frac \Lambda M =0.347$, and $M=312$~TeV for I.9, and $\kappa_3=0.908$, $N=3$, $\frac \Lambda M =0.290$, and $M=350$~TeV for I.13. The spectra are shown in figure \ref{fig:spectra}. 

\item Despite our focus on CP conserving observables in this work, there could, in principle, be significant constraints from $\epsilon_K$ introduced solely through the CP violating  phase of the CKM.\footnote{We thank W.~Altmannshofer for bringing this to our attention.}   In order to avoid this constraint, in our $Q$-class models, we choose $\kappa_1$, $\kappa_2$ and $\kappa_3$  to align with the low-energy down, strange and bottom quark, respectively. Thus in our choice of interaction basis,  the down Yukawa is diagonal, but the up Yukawa is multiplied by $V_{CKM}$.  We ignore the small differences  with  our previous work where the alignment was with the top quark.  The $U$-class models retain alignment with the up-type quarks.

\item Turning on $\kappa_1$ and $\kappa_2$ while holding fixed the other parameters can significantly modify the spectrum, as shown in (\ref{eq:QgenK}). At the least finely-tuned points, where a cancellation results in a stop lighter than the other squarks, this can either lead to stop tachyons, or it can lead to the stops  being so heavy that the hypercharge tadpole contribution to the RG running quickly drives the sleptons tachyonic.   To avoid these undesirable features, as we turn on $\kappa_{1,2}$, we fix $\Lambda$, but vary $M$, so that the the one-loop contribution adapts in such a way that  the lightest squark eigenvalue of the LL block in $Q$-class or RR block in $U$-class models is held fixed. This deformation enables us to prevent the interesting flavor-violation from vanishing and to maintain the squark masses at sensible values.  

\item Additionally, as can be seen from (\ref{eq:QgenK}), if we were to fix $\kappa_3$ and turn on $\kappa_{1,2}$, then the  ``net''  $A$-term would increase.  The ``net'' $A$-term in the type I models is aligned with the lightest eigenvalue direction of the rank 1 block, i.e., in $Q$-class models, 
\beq
\mathcal L \supset - \frac{\kappa \kappa_3 y_t \Lambda}{16\pi^2} H_u \tilde Q_S \tilde t_R.
\eeq 
The $A$-terms in all orthogonal squark directions vanish in the third-generation dominant limit.  Increasing $A/m_S$ is potentially dangerous for vacuum stability \cite{Park:2010wf}.  In order to avoid these issues, we require that this ``net'' $A$-term remains constant.  To achieve this we fix $\kappa_3 \kappa$ everywhere in the parameter space so that as we increase $\kappa_{1}$ and $\kappa_2$, $\kappa_3$ decreases to compensate.

\item Finally, this deformation of $\kappa_{1,2,3}$   will also affect the Higgs mass. The one-loop corrections to the Higgs mass in the presence of general flavor violation have been computed in \cite{Arana-Catania:2014ooa,Kowalska:2014opa}, and our preliminary studies of these suggest that  $m_h$ may drift down by several GeV as we move out in $\kappa_{1,2}$. However, the two-loop corrections are not yet known, and if the usual non-flavor-violating MSSM is any guide, these are likely to be important for an accurate determination of the Higgs mass. While it would be interesting to study this further, it is beyond the scope of the work.  At the very least, one could imagine increasing the overall scale of the superpartners in order to compensate for any decrease in the Higgs mass. This would only serve to further alleviate the flavor constraints, so the qualitative value of \chifv\ to these EGMSB models is unaffected.

\begin{figure}[!t]
\begin{center}
\includegraphics[scale=0.75]{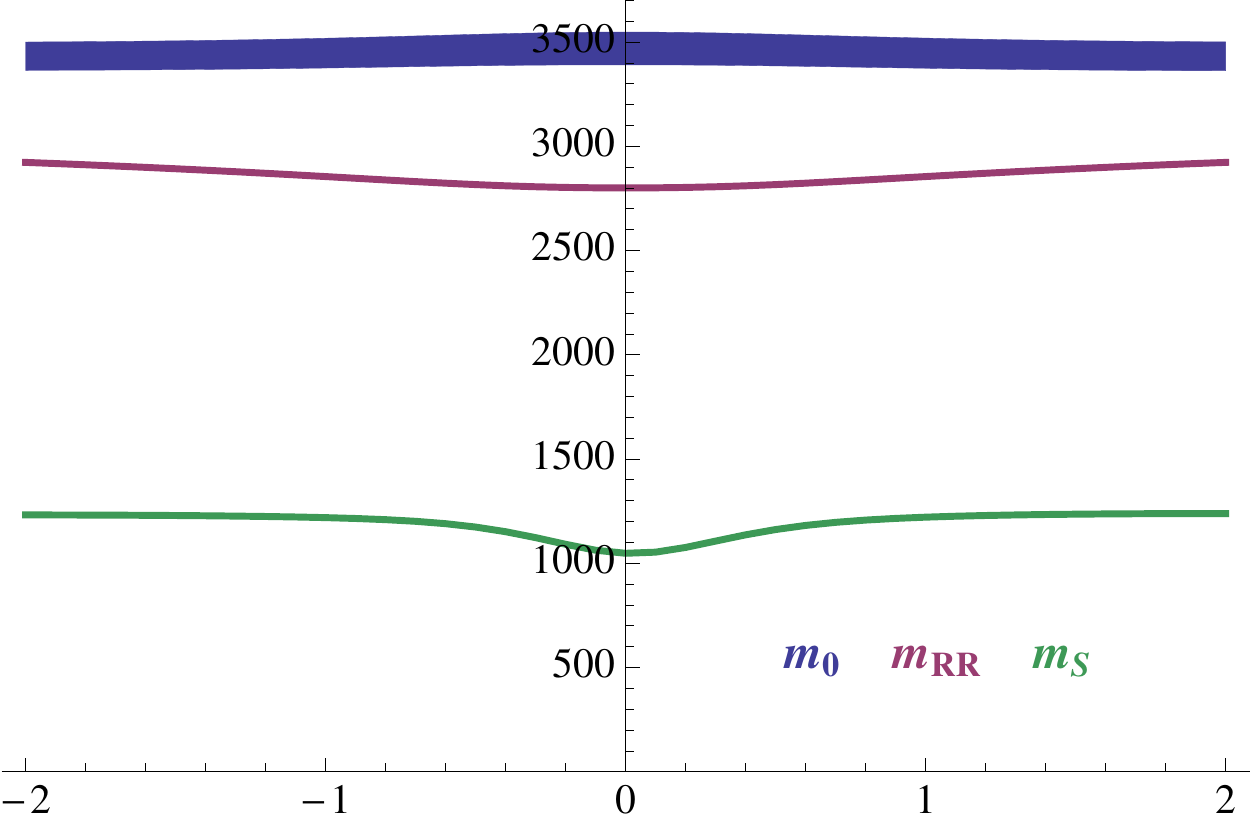}
\end{center}
\begin{picture}(0,0)(0,0)
\put(185,15){$\kappa_1/\kappa_3=\kappa_2/\kappa_3$}
\end{picture}
\caption{
 The spectra of squark masses (in GeV) for model I.9 as we move in $\kappa_1=\kappa_2$
Aside from the three lightest eigenvalues, $m_{S, up}=m_{S,down}$, and $m_{RR}$, the other nine squark eigenvalues all fall in the range of the blue band near 3.5 TeV.  The width of the band is largely due to the SU(2) GMSB contributions to the left-handed squarks.}
\label{fig:masses}
\end{figure}
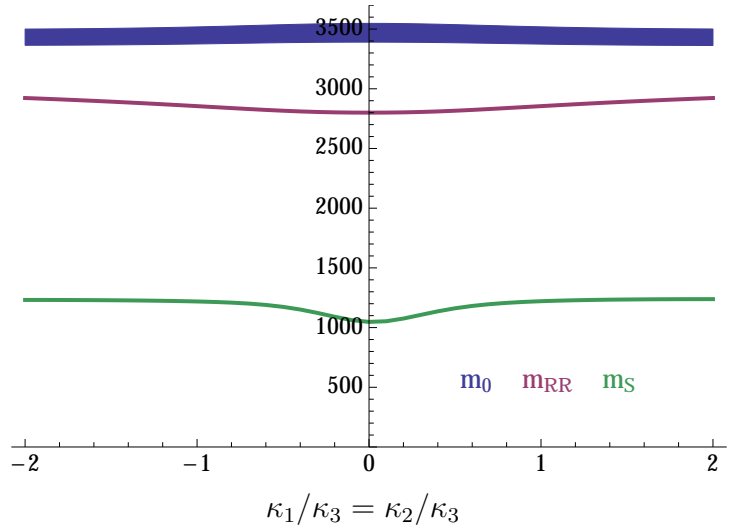

\end{itemize}

Across a grid in $(\kappa_1, \kappa_2)$ constructed via this deformation of the benchmark point, we generate the soft spectrum at the messenger scale. All of the couplings and soft masses are then evolved down from the messenger scale to the SUSY scale using the fully general $3\times 3$ MSSM $\beta$-functions.\footnote{For simplicity, we do not run back and forth between the IR and UV to better specify the scale.} At the SUSY scale, we apply the BMPZ QCD threshold corrections to the squarks and gluinos \cite{Pierce:1996zz}.  Finally, we  use our new $\mathtt{Mathematica}$ package \FormFlavor\ to compute the Wilson coefficients, RG evolve these coefficients down to the scale of interest (e.g., $m_b$) for each flavor observable, and compute the contributions to each flavor observable there.  (The details of \FormFlavor\ are briefly explained in appendix~\ref{app:FF} and will be further fleshed out when the package is made publicly available \cite{FFpaper}.)

\section{Details of \FormFlavor}
\label{app:FF}

\FormFlavor\ is a general, modular flavor package written in $\mathtt{Mathematica}$ that takes general MSSM spectra as inputs and calculates a variety of flavor observables \textit{in situ} starting from the Feynman rules. The package is flexible -- new flavor observables can be straightforwardly added to it and, although currently implemented only for the MSSM, it can be extended to other models readily. Also, since processes are calculated in a uniform manner from first principles (rather than hardcoding formulas from the literature), the reliability of the code is greatly enhanced. 

In \FormFlavor, the process under consideration is specified in $\mathtt{FeynArts}$~\cite{Hahn:2000kx} and all the relevant topologies for the various sparticle mediators are generated there. These topologies are then converted into their respective amplitudes via $\mathtt{FormCalc}$~\cite{Hahn:1998yk}. The $\mathtt{FormCalc}$ output is analytically transformed into a particular Wilson operator basis for each process of interest, and the Wilson coefficients are extracted.  Having performed this once for an observable the analytic expressions may be written to a process file and used in subsequent runs. The Wilson coefficients may then be piped into functions that calculate the flavor observables.  

The main components of \FormFlavor\ are as follows: 
\begin{itemize}
\item Automated generation of one-loop diagrams for flavor processes (using $\mathtt{FeynArts}$)
\item Automated calculation of Wilson coefficients from Feynman diagrams (using $\mathtt{FormCalc}$)
\item Library of analytic one-loop integral functions
\item Routines for general MSSM spectrum input (SLHA2 \cite{Allanach:2008qq} compatible)
\item Routines for converting Wilson coefficients to flavor observables. (This step contains hardcoded formulas from the literature and is not MSSM specific.) The following processes are currently implemented: \vspace{2mm}\\ \textit{ CP-conserving observables}
\begin{itemize}
\item[*] Meson Mixings \cite{Buras:2001ra,Aoki:2013ldr}: $\Delta M_K$ \cite{Bae:2013mja}, $\Delta M_D$ \cite{Carrasco:2013jaa}, $\Delta M_{B_d}$ and $\Delta M_{B_s}$ \cite{Carrasco:2013iba,Lenz:2014jya}
\item[*] Leptonic decays : $K^+\rightarrow \pi^+ \nu \bar\nu$ \cite{Buras:1998raa,Brod:2010hi,Cirigliano:2011ny}, $B_s \rightarrow \mu^+ \mu^-$ and $B_d \rightarrow \mu^+ \mu^-$ \cite{Bobeth:2002ch,Dedes:2008iw,Hermann:2013kca,Bobeth:2013uxa}
\item[*] Radiative Decays : $b\rightarrow s \gamma$ \cite{Buras:2002tp,Misiak:2006ab}, $b\rightarrow d \gamma$ \cite{Hurth:2003dk,Crivellin:2011ba}
\end{itemize}
\textit{CP-violating observables}
\begin{itemize}
\item[*] Meson Mixings : $\epsilon_K$
\item[*] Leptonic decays : $K^0\rightarrow \pi^0 \nu \bar\nu$
\item[*] Electric dipole moments : neutron-EDM
\end{itemize}
\item MSSM RGE -- Full three-family MSSM renormalization group evolution, including arbitrary CP-phases, implemented with the aid of $\mathtt{SARAH}$~\cite{Staub:2010jh} 
\end{itemize}
This powerful flavor package will be released to the public and described in more detail in an upcoming work \cite{FFpaper}.

\section{Soft Terms for Type I Squark Models }
\label{app:EGMSB}
\label{app:softTypeI}

In this appendix, we will specialize the fully general EGMSB formulas from \cite{Evans:2013kxa} to the case of the type I squark models with general flavor-violating couplings $\kappa_i$. These were quoted in (\ref{eq:QgenK}) for the $Q$-class models, 
\beq
\label{eq:supapp}
W_{EGMSB} =\kappa_{i} Q_i \sum_{A=1}^N \Phi_{A}\tilde\Phi_{A} 
\eeq
and we again focus on this case here. To translate to $U$-class models, one must take $U\leftrightarrow Q$, $y_u\to y_u^\dagger$, $A_{\tilde u}\to A_{\tilde u}^\dagger$ and $K\to K^T$ in all expressions.

Before we begin, we must address a convention difference between this paper and that of  \cite{Evans:2013kxa} concerning the treatment of the right-handed quark chiral superfields $\bar U$, $\bar D$. In this paper, the Yukawas, squark mass matrices and $A$-terms  are defined as:
\bea
W \supset& \;  H_u \bar U_i y_{u,ij} Q_j + H_d \bar D_i y_{d,ij} Q_j  \\ 
\mathcal L \supset& \;\; \tilde Q_i^*  \delta m_{Q,ij}^2 \tilde Q_j  + \tilde{ U}_i^* \delta m_{U,ij}^2 \tilde{  U}_j + \tilde{  D}_i^* \delta m_{D,ij}^2 \tilde{D}_j  \\& + \lp H_u \tilde{ U}_i^* A_{\tilde u,ij} \tilde Q_j + H_d \tilde{  D}_i^* A_{\tilde d,ij} \tilde Q_j +\hc\rp
\eea
Thus, the Yukawas, $A$-terms, $m_{U}^2$ and $m_D^2$ are all transposed relative to those of \cite{Evans:2013kxa}.   Note that in these conventions, $\tilde U^*$, $\tilde D^*$ are the scalar components of the {\it chiral} supermultipliets $\bar U$, $\bar D$. 

The starting point of our derivation of (\ref{eq:QgenK}) is eq.~(2.19) in \cite{Evans:2013kxa}, which we repeat here for convenience:
\bea
\label{eq:typeIgen}
A_{ab} &=  -{1\over32\pi^2}d_{a}^{BC} \lambda_{aBC}^*\lambda_{bBC} \Lambda\\
\delta m_{ab}^2 &= {1\over 256\pi^4}\Bigg( d_a^{ B C}d_{B}^{c D}\lambda_{a B C}^*\lambda_{b C E} \lambda_{c B D}\lambda_{c DE}^* + {1\over4}d_{a}^{B C }d_{b}^{DE} \lambda_{a B C}^* \lambda_{c B C } \lambda_{c DE}^* \lambda_{bDE}\\
&\qquad - {1\over2}d_a^{cd}d_c^{BC}y_{acd}^* y_{b d e}\lambda_{c BC}\lambda_{e BC}^* -d_{a}^{BC}C_r^{aBC}g_r^2  \lambda_{aBC}^*\lambda_{bBC}\Bigg)\Lambda^2
\eea
Here $a$, $b$, \dots run over all MSSM fields (including all gauge and flavor degrees of freedom). Meanwhile $A$, $B$, \dots run over all messenger fields similarly. In this appendix, we will use $i$, $j$, \dots to denote MSSM flavor indices.

From (\ref{eq:typeIgen}), we immediately obtain the bilinear $A$-terms after summing over the $N$ messenger multiplets and substituting $\lambda_{aBC}\to \kappa_i$ and $d_a^{BC}\to d_Q\propto N$
\beq
\label{eq:Aterms}
A_{Q,ij}  = -{1\over 16\pi^2} d_Q \kappa_i^*\kappa_j \Lambda
\eeq
This becomes the trilinear $A$-term $A_{\tilde u}$ used in the paper via:
\beq 
A_{Q,ij}F^\dagger_{Q_i} \tilde Q_j \to y_{u,ik}A_{Q,kj}  H_u \tilde U_i^*  \tilde Q_j  \equiv A_{\tilde u,ij} H_u \tilde U_i^* \tilde Q_j
\eeq

Next let's consider the EGMSB contributions to the soft mass-squareds, starting with $m_Q^2$. From (\ref{eq:typeIgen}), we obtain:
\bea
\label{eq:softmass1app}
\delta m_{Q,ij}^2 &= {1\over 256\pi^4}\Bigg( d_Q d_\phi \kappa_i^*\kappa_j \kappa_k \kappa_k^* + d_Q^2 \kappa_i^*  \kappa_k^* \kappa_k \kappa_j - 2 d_{Q} C_r g_r^2   \kappa_i^*\kappa_j\Bigg)\Lambda^2\\
 &= {d_Q\over 256\pi^4}\Bigg( (d_\phi + d_Q) \kappa^2 - 2 C_r g_r^2 \Bigg)K_{ij}\Lambda^2
\eea
In the first line, we have introduced $d_{B}^{cD}\to {1\over2}d_\phi$, and we have used the fact that $Q$ is the only MSSM field coupling to the messengers to set $d_a^{cd}d_c^{BC}$ to zero. Additionally, as $Q$ couples directly to the messengers, there is a one-loop term suppressed by $\frac {\Lambda^2}{M^2}$ \cite{Craig:2012xp},
\beq
\delta \lp m^{1\mbox{\scriptsize-loop}}_{Q,ij}\rp^2 = - \frac{16\pi^2}{3} h\lp\!\frac\Lambda M\!\rp \frac{\Lambda^2}{M^2} \frac{d_Q  K_{ij} \Lambda^2}{256\pi^4}
\eeq
where $h(x)$ is a loop function  given by 
\beq
h(x)=\frac 3{x^4}\bigg((x-2)\ln(1-x)-(x+2)\ln(1+x)\bigg)=1+\frac45 x^2 +\order{x^4}.
\label{eq:hfunc}
\eeq

The EGMSB contribution to $m_{H_u}^2$ is much simpler. Here only the third term of $\delta m_{ab}^2$ in (\ref{eq:typeIgen}) contributes:
\beq
\label{eq:softmass2app}
\delta m_{H_u}^2  =  - { 3 d_Q \over 256\pi^4}  y^{*}_{u,ij} y_{u,ik} \kappa_j \kappa_k^* \Lambda^2  =  - {3 d_Q \over 256\pi^4}\Tr\left[ y_u K y_{u}^\dagger \right] \Lambda^2
\eeq
where we have used $d_H^{QU}=3$. 

Lastly, the EGMSB contribution to $m_{U}^2$ also comes from just the third term of $\delta m_{ab}^2$:
\bea
\label{eq:softmass3app}
\delta m_{U,ij}^2 = - { d_U^{Q H} d_Q  \over 256\pi^4}  y^{*}_{u,ik} y_{j\ell}\kappa_k \kappa_\ell^* \Lambda^2=  - { d_U^{Q H} d_Q  \over 256\pi^4}  \left(y_u K y_{u}^\dagger \right)_{ji}  \Lambda^2
\eea
where $d_U^{QH}=2$. Taking into account the need to transpose $m_U^2$ to translate between the conventions of \cite{Evans:2013kxa} and those of this paper, we obtain the correct result quoted in (\ref{eq:QgenK}).

\section{Formulas for flavor observables}
\label{app:flavobs}

In this appendix, we collect formulas from the literature for the various flavor observables considered in this work. Along the way, we will streamline the different notations scattered throughout the literature into a uniform convention.

The uniform operator basis we will use was introduced in the text; we repeat it here for convenience.  For dimension 5, we have:  
\begin{eqnarray}
\label{eq:wilsonOpA}
& & \CO_A^M(f_1,f_2) = e \bar f_1 \sigma^{\mu\nu}P_M f_2 F_{\mu\nu}\\
& & \CO_G^M(f_1,f_2) = g \bar f_1 \sigma^{\mu\nu}P_M f_2 G_{\mu\nu}
\end{eqnarray}
For dimension 6, we have:
\begin{eqnarray}
& &\CO^{MN}_S(f_1,f_2,f_3,f_4) = (\bar f_1 P_M f_2) (\bar f_3 P_N f_4)\\
& & \CO^{MN}_V(f_1,f_2,f_3,f_4) = (\bar f_1 \gamma^\mu P_M f_2) (\bar f_3 \gamma_\mu P_N f_4)\\
& & \CO^{MN}_T(f_1,f_2,f_3,f_4) = (\bar f_1 \sigma^{\mu\nu} P_M f_2) (\bar f_3 \sigma_{\mu\nu} P_N f_4)
\end{eqnarray}
where $M,N=L,R$. $P_{R} = \frac 12 \lp 1 +\gamma_5\rp $, $P_{L} = \frac 12 \lp 1 -\gamma_5\rp $ are projection operators, $\sigma^{\mu\nu} = \frac 12 [\gamma^\mu,\gamma^\nu]$ and if $f_i$ carry color indices, they are contracted within a bilinear factor.

The general effective Hamiltonian is then:
\begin{equation}
H_{eff}(f_i) =\sum  C_X^M(f_i) \CO_X^M(f_i) +  \sum C_X^{MN}(f_i) \CO_X^{MN}(f_i)
\end{equation}
where the sums runs over a complete basis of independent operators. 

\subsection{Meson Mixing $\Delta m_X$}
\label{app:Dmx}

\begin{table}[htdp]
\begin{center}
\small
\begin{tabular}{|c|ccc|ccccc|c|}
\hline
\!Meson\! & \!\!$m_X$\footnotesize(GeV)\normalsize\!\! & \!\!$f_X$\footnotesize(GeV)\normalsize\!\! & $R_X$ & $B^{LL}_V$ & $B_{V}^{LR} $ & $B_{S}^{LR} $ & $B^{LL}_{S} $ & $B^{LL}_{T}$  & $C^{LL}_{V,SM}|_{\mu=m_b}$(GeV$^{-2}$)\normalsize  \\
\hline 
$\Delta m_K$ & $0.4976$ & $0.160$ & $24.3$ & $0.56$ & $0.85 $ & $1.08$ & $0.62$ & $0.43$ & $-$ \\
$\Delta m_D$ &  $1.8645$ & $0.209$ & $3.20$ & $0.76$ & $0.97$ & $0.95$ & $0.64$ & $0.39$ &  $-$    \\ 
$\Delta m_{B_d}$ & $5.2796$ & $0.191$ & $1.65$ &  $0.84$ & $1.47 $ & $0.95$ & $0.72$ & $0.61$ &  $(2.34-2.20i) \!\times\! 10^{-12}$  \\
$\Delta m_{B_s}$ & $5.3668$ & $0.228$ & $1.65$ &  $0.88$ & $1.57 $ & $0.93$ & $0.73$ & $0.62$ &  $(6.96-0.26i) \!\times\! 10^{-11}$ \\
\hline
\end{tabular}
\normalsize
\caption{Properties of the meson mixing used in this work.  $f_X$ and $B_{V,X}^{LL}$ come from the FLAG review \cite{Aoki:2013ldr}.  The other four non-perturbative $B$-parameters are taken from several sources: for $\Delta m_K$ from \cite{Bae:2013mja} (at $\mu=2$ GeV), $\Delta m_D$ from \cite{Carrasco:2013jaa} (at $\mu=3$ GeV, rescaled for a common $R_D$, and converted to our basis using $B^{LL}_T=\frac 53 B_2-\frac 23 B_3$), and for $B_s$ and $B_d$ from \cite{Carrasco:2013iba} (at $\mu=m_b=4.2$ GeV and converted to our basis).  $C^{LL}_{V,SM}|_{\mu=m_b}$ are the \FormFlavor\ values in the CKM basis of the PDG \cite{Beringer:1900zz}. }
\end{center}
\label{tab:MesonProperties}
\end{table}

This corresponds to a $\Delta F=2$ effective Hamiltonian, with $f_1=f_3=q_1$, and $f_2=f_4=q_2$, with $(q_1,q_2)=(s,d),(c,u),(b,d),(b,s)$ for $K$, $D$, $B_d$ and $B_s$ respectively. The quantities relevant for mixing are derived from the effective Hamiltonian as
\bea
\langle \bar X | H_{eff} | X\rangle & \equiv M_{X,12} - \frac{i}2\Gamma_{X,12}\\
\langle X | H_{eff} | \bar X\rangle &= \, M^*_{X,12} - \frac{i}2\Gamma^*_{X,12}
 \label{eq:mXHeff}
\eea
$M_{X,12}$ and $\Gamma_{X,12}$ are, respectively, the dispersive and absorptive parts of $\langle \bar X | H_{eff} | X\rangle$. In terms of these quantities, the mass splitting is given by:
\beq
\Delta m_X= 2 \Re \left[ \lp M_{X,12} - \frac{i}2\Gamma_{X,12}\rp \sqrt{\frac{M^*_{X,12} - \frac{i}2\Gamma^*_{X,12}}{M_{X,12} - \frac{i}2\Gamma_{X,12}}} \right]
\label{DeltamXgen}\eeq
For $X=B_d$ and $B_s$ where $\Gamma_{X,12}\ll M_{X,12}$, this is well approximated by
\beq
\label{DeltamB}
\Delta m_X\approx 2|M_{X,12}| \approx 2|\langle \bar X | H_{eff} | X\rangle|
\eeq
For $X=K$ and $D$, where experimentally one finds that $M_{X,12}/\Gamma_{X,12}$ is approximately real, and  $M_{X,12}$ and $\Gamma_{X,12}$ are both predicted to be approximately real in the standard CKM convention (where the CPV phase is primarily in $V_{td}$ and $V_{ub}$), one has to a good approximation
\beq
\label{DeltamKD}
\Delta m_X\approx 2\,{\rm Re}\,M_{X,12} \approx 2\,{\rm Re}\,\langle\bar X | H_{eff} | X\rangle
\eeq
Finally, the short-distance part of the matrix elements in (\ref{DeltamB}) and (\ref{DeltamKD}) (which is all that is relevant for the $B_q$ systems and for new physics) are given by
\bea
\langle\bar X | H_{eff} | X\rangle &=\, \frac{m_X f_X^2}{24} \Bigg(
 8 B_V^{LL} (C_{V}^{LL}  + C_{V}^{RR}) -
 R_X \Bigg[ 4 B_{V}^{LR} C_{V}^{LR}  - 6 B_{S}^{LR} C_{S}^{LR} \\
 &\qquad  + 5 B_{S}^{LL} \lp C_{S}^{LL} + C_{S}^{RR}\rp + 12 B_{T}^{LL} \lp C_{T}^{LL} + C_{T}^{RR}\rp\Bigg] \Bigg)
\eea
where,
\beq
R_X = \lp \frac{m_X}{m_{q_1}+m_{q_2}}\rp^2,
\eeq
and the $B$-parameters are non-perturbative corrections that have been computed on the lattice.   In the Standard Model, only $C^{LL}_{V}\neq 0$.  These values and the other parameters relevant for meson mixing are shown in Table \ref{tab:MesonProperties}.

\subsection{$K^\pm\rightarrow\pi^\pm\nu\bar\nu$} 
\label{app:Kpvv}

This corresponds to a dimension 6 effective Hamiltonian with $f_1=s$, $f_2=d$, $f_3=f_4=\nu_\ell$.  The branching ratios for $K^\pm\rightarrow\pi^\pm\nu\bar\nu$ are given by \cite{Buras:1998raa}, 
\bea
\label{eq:Kpinunucomp}
\text{BR}(K^\pm\rightarrow\pi^\pm\nu\bar\nu)=&\;\frac{ c_+}{3} v^4 \sum_{\ell=e,\mu,\tau}\left| {C}^{LL}_{V, \ell}+ {C}^{RL}_{V,\ell}\right|^2  \\
\eea
where $v=246 \gev$ and 
\bea
 c_+ &= \; \frac{3 \, r_{K^+} }{2 \abs{ V_{us}}^2} \, \text{BR}(K^\pm\rightarrow \pi^0 e^\pm \nu) = 1.35.
\eea
Here, the branching ratio has been included to remove dependence on the hadronic matrix element, and $r_{K^+}=0.901$ contains isospin violating quark mass effects and electroweak corrections computed in \cite{Marciano:1996wy}.  

In the standard model, the contributions from top loops are have no sensitivity to different generations, but the charm loop contributions do.  In particular, the charm contributions are the same for $e$ and $\mu$, but differ for $\tau$.   Thus, the SM contribution can be expressed as,
\beq
{C}^{LL}_{V,SM,\ell} =  \frac{\alpha_2 }{\pi v^2} \lp \lambda_c X^\ell_c + \lambda_t  X_t \rp 
\eeq
where $\lambda_i =  V^*_{is}V_{id}$, $X_t =1.469$ \cite{Brod:2010hi}, $X^e_c=X^\mu_c=1.055\times10^{-3}$, and $X^\tau_c=7.01\times10^{-4}$  \cite{Buras:1998raa}.

In the absence of lepton flavor-violating new physics effects, the contributions from new physics are the same across lepton generations, i.e., $C^{XY}_{V,NP,\ell}=C^{XY}_{V,NP,\ell'}\equiv C^{XY}_{V,NP}$.  We can then give a simpler form to equation (\ref{eq:Kpinunucomp})
\bea
\text{BR}(K^\pm\rightarrow\pi^\pm\nu\bar\nu)=&\;  c_+ v^4 \left| {C}^{LL}_V+ {C}^{RL}_V\right|^2  \\
\eea
where $C^{XY}_V=C^{XY}_{V,NP}+C^{XY}_{V,SM}$, with 
\bea
{C}^{RL}_{V,SM} &=\; 0 \\
{C}^{LL}_{V,SM} &=\;  \frac{\alpha_2 }{\pi v^2} \lp \lambda_c P_c + \lambda_t  X_t \rp =(-12.2+3.6 i)\times 10^{-11} \gev^{-2} \\
P_c &= \; \lp \frac23 X^e_{NL} +\frac13 X^\tau_{NL} \rp \sim 9.37 \times 10^{-4}.
\eea
Here, $P_c$ is the charm contribution averaged over the different neutrino flavors.  Using $P_c$ simplifies the expression, but reduces the standard model charm contribution by about 3\%.  This represents only a 0.3\% decrease in the overall SM contribution, which is completely negligible when compared to the theoretical uncertainty.  Most importantly, the interference effects with new physics are properly captured under this simplification.

\subsection{$b\rightarrow s \gamma$ and $b\rightarrow d \gamma$} 
\label{app:bsg}

These observables correspond to dimension 5 effective Hamiltonians with $f_1=b$ and $f_2=s$ or $d$. The branching ratio is given by,
\be
\text{BR}(b\rightarrow q \gamma)=  c_\gamma v^2 \Big( \abs{  C_A^L}^2 + \abs{ C_A^R}^2 \Big)
\ee
where \cite{Bauer:2004ve}
\bea
 c_\gamma  &=\; \lp8 \pi^2 \rp^2\frac{6}{\pi} \frac{\text{BR}(b\rightarrow X_c e \nu)_{\text{\tiny{EXP}}}}{\Phi \abs{V_{cb} }^2} \frac{v^2}{m_b^2}\alpha_{\text{\tiny{EM}}} =  3.3 \times 10^{7}\\
\Phi &=\; \abs{\frac{V_{ub}}{V_{cb}}}^2 \frac{\text{BR}(b\rightarrow X_c e \nu)}{\text{BR}(b\rightarrow X_u e \nu)} = 0.58 
\eea
where the branching ratio to charm decays is used  to remove sensitivity to the hadronic matrix element, and the $\Phi$ factor is introduced to account for the nontrivial phase space factor in the compared branching ratio (due mostly to the charm quark mass).

 Only $C_A^L$ contains a standard model contribution, which is,
\beq
\left[C^L_{A,SM}\right]_{b\to s \gamma} = \frac{ V_{ts}^*V_{tb} m_b}{8\pi^2 v^2} X_{SM} \sim \lp 1.3 + 0.03 i \rp \times10^{-8}\gev^{-1} 
\eeq
where $X_{SM}=-0.3736$.

The observable $b\to d \gamma$ is defined completely analogously.  The only difference is that the the standard model contribution differs.  In particular, 
\beq
\left[C^L_{A,SM}\right]_{b\to d\gamma} = \frac{ V_{td}^*V_{tb}m_b}{8\pi^2 v^2} X_{SM} \sim -\lp 2.4 + 1.1 i \rp \times10^{-9}\gev^{-1} 
\eeq
with the same $X_{SM}=-0.3736$ as above.

\subsection{$B_s\rightarrow \mu^+ \mu^-$ and $B_d\rightarrow \mu^+ \mu^-$}
\label{app:bmm}

This corresponds to a dimension 6 effective Hamiltonian with $f_1=b$, $f_2=s,d$ and $f_3=f_4=\mu$. The branching ratio for $B_{s,d}\to\mu^+\mu^-$ is \cite{Bobeth:2002ch,Dedes:2008iw}:
\begin{equation}
\text{BR}(B_i\rightarrow \mu^+ \mu^-) = X_i \left\{  \left(1-\frac{4 m_\mu^2}{m_{B_i}^2}\right) | {F}^{(i)}_S |^2 + | {F}^{(i)}_P + {F}^{(i)}_A |^2\right\}
\end{equation}
where 
\beq
X_i=\frac{f_{B_i}^2 }{128 \pi \,m_{B_i} \Gamma_{B_i}} \sqrt{1-\frac{4 m_\mu^2}{m_{B_i}^2}}  \; \Longrightarrow \; X_s=5.36\times 10^7\mbox{ and } X_d=3.97\times 10^7,
\eeq
\bea
\label{eq:bmmFdefapp}
{F}^{(i)}_S =&\;  {m_{B_i}^3\over m_b+m_i} ({ C}_S^{ LL}+{ C}_S^{ LR}-{ C}_S^{ RR}-{ C}_S^{ RL}), \\ 
{F}^{(i)}_P =&\;  {m_{B_i}^3\over m_b+m_i}  (-{ C}_S^{ LL}+{ C}_S^{ LR}-{ C}_S^{ RR}+{ C}_S^{ RL}), \\ 
{F}^{(i)}_A =&\; 2m_{B_i}m_\mu ({ C}_V^{ LL}-{ C}_V^{ LR}+{ C}_V^{ RR}-{ C}_V^{ RL}),
\eea
and $\Gamma_{B_d}\equiv \tau_{B_d}^{-1} =4.33\times 10^{-13}$ GeV and $\Gamma_{B_s}\equiv \tau_{B_s}^{-1}=4.49\times 10^{-13}$ GeV.  The three-loop Standard Model contribution to $B_i\to \mu^+\mu^-$ can be expressed as \cite{Hermann:2013kca}:
\beq
F^{(i)}_{A,SM} = 0.4802\, \times \frac{4 \alpha_2 V_{tb}^*V_{ti}}{\pi}  \frac{m_\mu m_{B_i}}{v^2}
 ; \; F^{(i)}_{S,SM}=F^{(i)}_{P,SM}=0
\eeq
so that $F^{(d)}_{A,SM}=(1.5-0.6 i)\times 10^{-9}$ and $F^{(s)}_{A,SM}=(-7.9-0.1 i)\times 10^{-9}$

\section{Rank 1 $\chi$FV Loop Functions}
\label{app:Loops}
We compile some of the loop functions for our rank 1 \chifv\ approximation below.

Starting with meson-mixing, we have,
\bea
f_{\tilde{g}}^{\Delta M,\text{\tiny{box}}}(x_g, x _{q})=\;&\frac{  2 x_q \log x _{q} \big(11 x_{q}^3 + 6 x_g x_{q}^2-2 x_g^2 x_{q}-13x_g x_{q}-2 x_g^2\big) }{ \big(x _{q}-1\big) \big(x_q-x_{g}\big)^3}\\
+&\frac{ 2 x_q x_g \log x_g \big(x _{q}-1\big)^2 \big(2 x_q x_g+13 x_q-17 x_g^2+2 x_g\big)}{\big(x_g-1\big)^3 \big(x_q-x_{g}\big)^3}\\
-&\frac{x_q\lp19 x_q^2 x_g+11 x_q^2+3 x_q x_g^2-74 x_q x_g+11 x_q+8 x_g^3+3 x_g^2+19 x_g\rp}{\big(x_g-1\big)^2 \big(x_q-x_{g}\big)^2}
\eea
Here, $x _{q}=m_{S}^2/m_{0}^2$ and $x_g=m_{\tilde{g}}^2/m_{0}^2$; where $m_{0}$ and $m_{S}$ are the heaviest and lightest squark mass eigenvalues.

For $K^\pm \rightarrow \pi^\pm \nu \bar{\nu}$, there are two contributions,
\beq
f_{\tilde{\chi}^\pm}^{K\rightarrow \pi \nu \nu}(x_\ell,x_2,x_\mu) = f_{\tilde{\chi}^\pm}^{K\rightarrow \pi \nu \nu,\text{\tiny{box}}}(x_\ell,x_2) + f_{\tilde{\chi}^\pm}^{K\rightarrow \pi \nu \nu,\text{\tiny{peng}}}(x_2,x_\mu).
\eeq
In practice, $f_{\tilde{\chi}^\pm}^{K\rightarrow \pi \nu \nu,\text{\tiny{peng}}}(x_\ell,x_2)$ is  numerically only $\order{10\%}$ of $f_{\tilde{\chi}^\pm}^{K\rightarrow \pi \nu \nu,\text{\tiny{box}}}(x_\ell,x_2)$, so, out of simplicity, we only present the box contribution for interpreting our results,
\bea
f_{\tilde{\chi}^\pm}^{K\rightarrow \pi \nu \nu,\text{\tiny{box}}}(x_\ell,x_2) =\;& 6  \left[ \frac{x_2\big(x_2^2-x_\ell\big) \log x_2}{\big(x_2-1\big){}^2 \big(x_2-x_\ell\big){}^2 }-\frac{x_2}{\big(x_2-1\big) \big(x_2-x_\ell\big)}
\right. \\
&   \hspace{45mm} \left.
-\frac{ x_2 x_\ell \log x_\ell }{\big(x_2-x_\ell\big){}^2 \big(x_\ell-1\big)} \right]
\eea
Here $x_l=m_{\tilde{l}}^2/m_{S}^2$, $x_2=M_2^2/m_{S}^2$ and $x_\mu=\mu^2/m_{S}^2$.  We have also dropped the neutralino boxes and $Z$-penguins that contribute at the $\order{10\%}$ level or below.

For $b\rightarrow q \gamma$
\bea
f_{\tilde{\chi}^\pm}^{ b \rightarrow s/d\,\gamma,\text{\tiny{peng}}}(x_\mu,x_2)=\;&  \frac{6 \sqrt{x_\mu x_2}}{11} \left[ \frac{5 x_2\, x_{\mu }-7 x_{\mu }-7 x_2+9}{ \big(x_2-1\big)^2 \big(x_{\mu }-1\big)^2} + \frac{ 2 \big(2 x_2-3\big) \log x_2}{ \big(x_2-1\big){}^3 \big(x_2-x_{\mu }\big)}
\right. \\
&   \hspace{55mm} \left.
- \frac{2\big(2 x_{\mu }-3\big) \log x_{\mu }}{ \big(x_2-x_{\mu }\big) \big(x_{\mu }-1\big){}^3} \right] 
\eea
Here $x_\mu= \mu^2/m_{S}^2$, $x_2=M_2^2/m_{S}^2$ and $m_{S}$ is the lightest up squark mass eigenvalue.

For $B_q\rightarrow \mu^- \mu^+$
\bea
f_{\tilde{g}}^{ B_q \rightarrow \mu^+ \mu^-,\text{\tiny{h-peng}}}(x_q,x_g)=&\sqrt{x_g x_q} \left[ \frac{\left(x_q-1\right) x_g \log x_g}{\left(x_g-1\right)^2 \left(x_g-x_q\right)} +\frac{1}{\left(x_g-1\right)} -\frac{ x_q \log x_q}{ \left(x_q-1\right) \left(x_g-x_q\right)}\right] 
\eea
\bea
f_{\tilde\chi}^{ B_q \rightarrow \mu^+ \mu^-,\text{\tiny{h-peng}}}(x_q,x_\mu,x_2)=& \sqrt{x_2} \left[ \frac{x_2 \big(x_q-1\big) \log x_2}{\big(x_2-1\big) \big(x_2-x_{\mu }\big) \big(x_2-x_q\big)}
\right. \\
&  \left.
+  \frac{x_{\mu } \big(x_q-1\big)  \log x_{\mu}}{\big(x_{\mu }-x_2\big) \big(x_{\mu }-1\big) \big(x_{\mu }-x_q\big)}
 +\frac{ x_q \log x_q}{\big(x_{\mu }-x_q\big)\big(x_2-x_q\big)}    \right]
\eea
Here $x_g= m_{\tilde{g}}^2/m_{0}^2$, $x_\mu= \mu^2/m_{0}^2$, $x_2=M_2^2/m_{0}^2$ and and $x _{q}=m_{S}^2/m_{0}^2$; $m_{S}$ and $m_{0}$ in this context are the lightest and heaviest squark mass eigenvalues.


\bibliography{atermbib}

\providecommand{\href}[2]{#2}\begingroup\raggedright\begin{thebibliography}{10}

\bibitem{Martin:1997ns}
S.~P. Martin, {\it {A Supersymmetry primer}},  {\em Adv.Ser.Direct.High Energy
  Phys.} {\bf 21} (2010) 1--153,
  [\href{http://arxiv.org/abs/hep-ph/9709356}{{\tt hep-ph/9709356}}].

\bibitem{Isidori:2010kg}
G.~Isidori, Y.~Nir, and G.~Perez, {\it {Flavor Physics Constraints for Physics
  Beyond the Standard Model}},  {\em Ann.Rev.Nucl.Part.Sci.} {\bf 60} (2010)
  355, [\href{http://arxiv.org/abs/1002.0900}{{\tt arXiv:1002.0900}}].

\bibitem{Giudice:1998bp}
G.~Giudice and R.~Rattazzi, {\it {Theories with gauge mediated supersymmetry
  breaking}},  {\em Phys.Rept.} {\bf 322} (1999) 419--499,
  [\href{http://arxiv.org/abs/hep-ph/9801271}{{\tt hep-ph/9801271}}].

\bibitem{Aad:2012tfa}
{\bf ATLAS} Collaboration, G.~Aad {\em et~al.}, {\it {Observation of a new
  particle in the search for the Standard Model Higgs boson with the ATLAS
  detector at the LHC}},  {\em Phys.Lett.} {\bf B716} (2012) 1--29,
  [\href{http://arxiv.org/abs/1207.7214}{{\tt arXiv:1207.7214}}].

\bibitem{Chatrchyan:2012ufa}
{\bf CMS} Collaboration, S.~Chatrchyan {\em et~al.}, {\it {Observation of a new
  boson at a mass of 125 GeV with the CMS experiment at the LHC}},  {\em
  Phys.Lett.} {\bf B716} (2012) 30--61,
  [\href{http://arxiv.org/abs/1207.7235}{{\tt arXiv:1207.7235}}].

\bibitem{Hall:2011aa}
L.~J. Hall, D.~Pinner, and J.~T. Ruderman, {\it {A Natural SUSY Higgs Near 126
  GeV}},  {\em JHEP} {\bf 1204} (2012) 131,
  [\href{http://arxiv.org/abs/1112.2703}{{\tt arXiv:1112.2703}}].

\bibitem{Heinemeyer:2011aa}
S.~Heinemeyer, O.~Stal, and G.~Weiglein, {\it {Interpreting the LHC Higgs
  Search Results in the MSSM}},  {\em Phys.Lett.} {\bf B710} (2012) 201--206,
  [\href{http://arxiv.org/abs/1112.3026}{{\tt arXiv:1112.3026}}].

\bibitem{Arbey:2011ab}
A.~Arbey, M.~Battaglia, A.~Djouadi, F.~Mahmoudi, and J.~Quevillon, {\it
  {Implications of a 125 GeV Higgs for supersymmetric models}},  {\em
  Phys.Lett.} {\bf B708} (2012) 162--169,
  [\href{http://arxiv.org/abs/1112.3028}{{\tt arXiv:1112.3028}}].

\bibitem{Draper:2011aa}
P.~Draper, P.~Meade, M.~Reece, and D.~Shih, {\it {Implications of a 125 GeV
  Higgs for the MSSM and Low-Scale SUSY Breaking}},  {\em Phys.Rev.} {\bf D85}
  (2012) 095007, [\href{http://arxiv.org/abs/1112.3068}{{\tt
  arXiv:1112.3068}}].

\bibitem{Carena:2011aa}
M.~Carena, S.~Gori, N.~R. Shah, and C.~E. Wagner, {\it {A 125 GeV SM-like Higgs
  in the MSSM and the $\gamma \gamma$ rate}},  {\em JHEP} {\bf 1203} (2012)
  014, [\href{http://arxiv.org/abs/1112.3336}{{\tt arXiv:1112.3336}}].

\bibitem{Evans:2012hg}
J.~L. Evans, M.~Ibe, S.~Shirai, and T.~T. Yanagida, {\it {A 125GeV Higgs Boson
  and Muon g-2 in More Generic Gauge Mediation}},  {\em Phys.Rev.} {\bf D85}
  (2012) 095004, [\href{http://arxiv.org/abs/1201.2611}{{\tt
  arXiv:1201.2611}}].

\bibitem{Kang:2012ra}
Z.~Kang, T.~Li, T.~Liu, C.~Tong, and J.~M. Yang, {\it {A Heavy SM-like Higgs
  and a Light Stop from Yukawa-Deflected Gauge Mediation}},  {\em Phys.Rev.}
  {\bf D86} (2012) 095020, [\href{http://arxiv.org/abs/1203.2336}{{\tt
  arXiv:1203.2336}}].

\bibitem{Craig:2012xp}
N.~Craig, S.~Knapen, D.~Shih, and Y.~Zhao, {\it {A Complete Model of Low-Scale
  Gauge Mediation}},  {\em JHEP} {\bf 1303} (2013) 154,
  [\href{http://arxiv.org/abs/1206.4086}{{\tt arXiv:1206.4086}}].

\bibitem{Albaid:2012qk}
A.~Albaid and K.~Babu, {\it {Higgs boson of mass 125 GeV in GMSB models with
  messenger-matter mixing}},  {\em Phys.Rev.} {\bf D88} (2013) 055007,
  [\href{http://arxiv.org/abs/1207.1014}{{\tt arXiv:1207.1014}}].

\bibitem{Abdullah:2012tq}
M.~Abdullah, I.~Galon, Y.~Shadmi, and Y.~Shirman, {\it {Flavored Gauge
  Mediation, A Heavy Higgs, and Supersymmetric Alignment}},  {\em JHEP} {\bf
  1306} (2013) 057, [\href{http://arxiv.org/abs/1209.4904}{{\tt
  arXiv:1209.4904}}].

\bibitem{Perez:2012mj}
M.~J. PŽrez, P.~Ramond, and J.~Zhang, {\it {Mixing supersymmetry and family
  symmetry breakings}},  {\em Phys.Rev.} {\bf D87} (2013), no.~3 035021,
  [\href{http://arxiv.org/abs/1209.6071}{{\tt arXiv:1209.6071}}].

\bibitem{Kim:2012vz}
H.~D. Kim, D.~Y. Mo, and M.-S. Seo, {\it {Neutrino Assisted Gauge Mediation}},
  {\em Eur.Phys.J.} {\bf C73} (2013), no.~6 2449,
  [\href{http://arxiv.org/abs/1211.6479}{{\tt arXiv:1211.6479}}].

\bibitem{Byakti:2013ti}
P.~Byakti and T.~S. Ray, {\it {Burgeoning the Higgs mass to 125 GeV through
  messenger-matter interactions in GMSB models}},  {\em JHEP} {\bf 1305} (2013)
  055, [\href{http://arxiv.org/abs/1301.7605}{{\tt arXiv:1301.7605}}].

\bibitem{Craig:2013wga}
N.~Craig, S.~Knapen, and D.~Shih, {\it {General Messenger Higgs Mediation}},
  {\em JHEP} {\bf 1308} (2013) 118, [\href{http://arxiv.org/abs/1302.2642}{{\tt
  arXiv:1302.2642}}].

\bibitem{Evans:2013kxa}
J.~A. Evans and D.~Shih, {\it {Surveying Extended GMSB Models with $m_{h}$=125
  GeV}},  {\em JHEP} {\bf 1308} (2013) 093,
  [\href{http://arxiv.org/abs/1303.0228}{{\tt arXiv:1303.0228}}].

\bibitem{Calibbi:2013mka}
L.~Calibbi, P.~Paradisi, and R.~Ziegler, {\it {Gauge Mediation beyond Minimal
  Flavor Violation}},  {\em JHEP} {\bf 1306} (2013) 052,
  [\href{http://arxiv.org/abs/1304.1453}{{\tt arXiv:1304.1453}}].

\bibitem{Fischler:2013tva}
W.~Fischler and W.~Tangarife, {\it {Vector-like Fields, Messenger Mixing and
  the Higgs mass in Gauge Mediation}},  {\em JHEP} {\bf 1405} (2014) 151,
  [\href{http://arxiv.org/abs/1310.6369}{{\tt arXiv:1310.6369}}].

\bibitem{Mummidi:2013hba}
V.~S. Mummidi and S.~K. Vempati, {\it {A little more Gauge Mediation and the
  light Higgs mass}},  {\em Nucl.Phys.} {\bf B881} (2014) 181--205,
  [\href{http://arxiv.org/abs/1311.4280}{{\tt arXiv:1311.4280}}].

\bibitem{Knapen:2013zla}
S.~Knapen and D.~Shih, {\it {Higgs Mediation with Strong Hidden Sector
  Dynamics}},  {\em JHEP} {\bf 1408} (2014) 136,
  [\href{http://arxiv.org/abs/1311.7107}{{\tt arXiv:1311.7107}}].

\bibitem{Ding:2013pya}
R.~Ding, T.~Li, F.~Staub, and B.~Zhu, {\it {Focus Point Supersymmetry in
  Extended Gauge Mediation}},  {\em JHEP} {\bf 1403} (2014) 130,
  [\href{http://arxiv.org/abs/1312.5407}{{\tt arXiv:1312.5407}}].

\bibitem{Liu:2013vaa}
C.~Liu and Z.-h. Zhao, {\it {A Realization of Effective SUSY with Strong
  Unification}},  {\em Phys.Rev.} {\bf D89} (2014), no.~5 057701,
  [\href{http://arxiv.org/abs/1312.7389}{{\tt arXiv:1312.7389}}].

\bibitem{Calibbi:2014pza}
L.~Calibbi, A.~Mariotti, C.~Petersson, and D.~Redigolo, {\it {Selectron NLSP in
  Gauge Mediation}},  {\em JHEP} {\bf 1409} (2014) 133,
  [\href{http://arxiv.org/abs/1405.4859}{{\tt arXiv:1405.4859}}].

\bibitem{Basirnia:2015vga}
A.~Basirnia, D.~Egana-Ugrinovic, S.~Knapen, and D.~Shih, {\it {125 GeV Higgs
  from Tree-Level $A$-terms}},  \href{http://arxiv.org/abs/1501.00997}{{\tt
  arXiv:1501.00997}}.

\bibitem{Chacko:2001km}
Z.~Chacko and E.~Ponton, {\it {Yukawa deflected gauge mediation}},  {\em
  Phys.Rev.} {\bf D66} (2002) 095004,
  [\href{http://arxiv.org/abs/hep-ph/0112190}{{\tt hep-ph/0112190}}].

\bibitem{Shadmi:2011hs}
Y.~Shadmi and P.~Z. Szabo, {\it {Flavored Gauge-Mediation}},  {\em JHEP} {\bf
  1206} (2012) 124, [\href{http://arxiv.org/abs/1103.0292}{{\tt
  arXiv:1103.0292}}].

\bibitem{Evans:2011bea}
J.~L. Evans, M.~Ibe, and T.~T. Yanagida, {\it {Relatively Heavy Higgs Boson in
  More Generic Gauge Mediation}},  {\em Phys.Lett.} {\bf B705} (2011) 342--348,
  [\href{http://arxiv.org/abs/1107.3006}{{\tt arXiv:1107.3006}}].

\bibitem{Galon:2013jba}
I.~Galon, G.~Perez, and Y.~Shadmi, {\it {Non-Degenerate Squarks from Flavored
  Gauge Mediation}},  {\em JHEP} {\bf 1309} (2013) 117,
  [\href{http://arxiv.org/abs/1306.6631}{{\tt arXiv:1306.6631}}].

\bibitem{Jelinski:2014uba}
T.~Jelinski and J.~Pawelczyk, {\it {Masses and FCNC in Flavoured GMSB scheme}},
   \href{http://arxiv.org/abs/1406.4001}{{\tt arXiv:1406.4001}}.

\bibitem{Calibbi:2014yha}
L.~Calibbi, P.~Paradisi, and R.~Ziegler, {\it {Lepton Flavor Violation in
  Flavored Gauge Mediation}},  {\em Eur.Phys.J.} {\bf C74} (2014), no.~12 3211,
  [\href{http://arxiv.org/abs/1408.0754}{{\tt arXiv:1408.0754}}].

\bibitem{CPVpaper}
J.~A. Evans and D.~Shih. {\it The SUSY CP Problem in EGMSB Models}, to appear.

\bibitem{Hahn:2000kx}
T.~Hahn, {\it {Generating Feynman diagrams and amplitudes with FeynArts 3}},
  {\em Comput.Phys.Commun.} {\bf 140} (2001) 418--431,
  [\href{http://arxiv.org/abs/hep-ph/0012260}{{\tt hep-ph/0012260}}].

\bibitem{Hahn:1998yk}
T.~Hahn and M.~Perez-Victoria, {\it {Automatized one loop calculations in
  four-dimensions and D-dimensions}},  {\em Comput.Phys.Commun.} {\bf 118}
  (1999) 153--165, [\href{http://arxiv.org/abs/hep-ph/9807565}{{\tt
  hep-ph/9807565}}].

\bibitem{Porod:2014xia}
W.~Porod, F.~Staub, and A.~Vicente, {\it {A Flavor Kit for BSM models}},  {\em
  Eur.Phys.J.} {\bf C74} (2014), no.~8 2992,
  [\href{http://arxiv.org/abs/1405.1434}{{\tt arXiv:1405.1434}}].

\bibitem{FFpaper}
J.~A. Evans and D.~Shih. {\it FormFlavor v1.0: Manual and Validation}, to
  appear.

\bibitem{Lenz:2014jya}
A.~Lenz, {\it {$B$-mixing in and beyond the Standard model}},
  \href{http://arxiv.org/abs/1409.6963}{{\tt arXiv:1409.6963}}.

\bibitem{Brod:2010hi}
J.~Brod, M.~Gorbahn, and E.~Stamou, {\it {Two-Loop Electroweak Corrections for
  the $K \to \pi \nu \bar{\nu}$ Decays}},  {\em Phys.Rev.} {\bf D83} (2011)
  034030, [\href{http://arxiv.org/abs/1009.0947}{{\tt arXiv:1009.0947}}].

\bibitem{Misiak:2006ab}
M.~Misiak and M.~Steinhauser, {\it {NNLO QCD corrections to the $\bar B\to X_s
  \gamma$ matrix elements using interpolation in $m_c$}},  {\em Nucl.Phys.}
  {\bf B764} (2007) 62--82, [\href{http://arxiv.org/abs/hep-ph/0609241}{{\tt
  hep-ph/0609241}}].

\bibitem{delAmoSanchez:2010ae}
{\bf BaBar} Collaboration, P.~del Amo~Sanchez {\em et~al.}, {\it {Study of
  $B\to X \gamma$ decays and determination of $\abs{V_{td}/V_{ts}}$}},  {\em
  Phys.Rev.} {\bf D82} (2010) 051101,
  [\href{http://arxiv.org/abs/1005.4087}{{\tt arXiv:1005.4087}}].

\bibitem{Crivellin:2011ba}
A.~Crivellin and L.~Mercolli, {\it {$B \to X_d \gamma$ and constraints on new
  physics}},  {\em Phys.Rev.} {\bf D84} (2011) 114005,
  [\href{http://arxiv.org/abs/1106.5499}{{\tt arXiv:1106.5499}}].

\bibitem{CMSandLHCbCollaborations:2013pla}
{\bf CMS and LHCb} Collaboration, CMS and LHCb, {\it {Combination of results on
  the rare decays $B^0_{(s)} \to \mu^+\mu^-$ from the CMS and LHCb
  experiments}}, .

\bibitem{Bobeth:2013uxa}
C.~Bobeth, M.~Gorbahn, T.~Hermann, M.~Misiak, E.~Stamou, {\em et~al.}, {\it
  {$B_{s,d} \to \ell^+ \ell^-$ in the Standard Model with Reduced Theoretical
  Uncertainty}},  {\em Phys.Rev.Lett.} {\bf 112} (2014) 101801,
  [\href{http://arxiv.org/abs/1311.0903}{{\tt arXiv:1311.0903}}].

\bibitem{Beringer:1900zz}
{\bf Particle Data Group} Collaboration, J.~Beringer {\em et~al.}, {\it {Review
  of Particle Physics (RPP)}},  {\em Phys.Rev.} {\bf D86} (2012) 010001.

\bibitem{Amhis:2012bh}
{\bf Heavy Flavor Averaging Group} Collaboration, Y.~Amhis {\em et~al.}, {\it
  {Averages of B-Hadron, C-Hadron, and tau-lepton properties as of early
  2012}},  \href{http://arxiv.org/abs/1207.1158}{{\tt arXiv:1207.1158}}.

\bibitem{Altmannshofer:2009ne}
W.~Altmannshofer, A.~J. Buras, S.~Gori, P.~Paradisi, and D.~M. Straub, {\it
  {Anatomy and Phenomenology of FCNC and CPV Effects in SUSY Theories}},  {\em
  Nucl.Phys.} {\bf B830} (2010) 17--94,
  [\href{http://arxiv.org/abs/0909.1333}{{\tt arXiv:0909.1333}}].

\bibitem{Altmannshofer:2013lfa}
W.~Altmannshofer, R.~Harnik, and J.~Zupan, {\it {Low Energy Probes of PeV Scale
  Sfermions}},  {\em JHEP} {\bf 1311} (2013) 202,
  [\href{http://arxiv.org/abs/1308.3653}{{\tt arXiv:1308.3653}}].

\bibitem{ChiFVpaper}
J.~A. Evans and D.~Shih. {\it Chiral Flavor Violation}, to appear.

\bibitem{Buras:2001ra}
A.~J. Buras, S.~Jager, and J.~Urban, {\it {Master formulae for Delta F=2 NLO
  QCD factors in the standard model and beyond}},  {\em Nucl.Phys.} {\bf B605}
  (2001) 600--624, [\href{http://arxiv.org/abs/hep-ph/0102316}{{\tt
  hep-ph/0102316}}].

\bibitem{Grinstein:1990tj}
B.~Grinstein, R.~P. Springer, and M.~B. Wise, {\it {Strong Interaction Effects
  in Weak Radiative $\bar{B}$ Meson Decay}},  {\em Nucl.Phys.} {\bf B339}
  (1990) 269--309.

\bibitem{Misiak:2015xwa}
M.~Misiak, H.~Asatrian, R.~Boughezal, M.~Czakon, T.~Ewerth, {\em et~al.}, {\it
  {Updated NNLO QCD predictions for the weak radiative B-meson decays}},
  \href{http://arxiv.org/abs/1503.01789}{{\tt arXiv:1503.01789}}.

\bibitem{Bobeth:2002ch}
C.~Bobeth, T.~Ewerth, F.~Kruger, and J.~Urban, {\it {Enhancement of
  $\mathcal{B}(\bar B_d \to \mu^{+} \mu^{-)}/ \mathcal{B}(\bar B_s \to \mu^{+}
  \mu^{-)}$ in the MSSM with minimal flavor violation and large tan $\beta$}},
  {\em Phys.Rev.} {\bf D66} (2002) 074021,
  [\href{http://arxiv.org/abs/hep-ph/0204225}{{\tt hep-ph/0204225}}].

\bibitem{Dedes:2008iw}
A.~Dedes, J.~Rosiek, and P.~Tanedo, {\it {Complete One-Loop MSSM Predictions
  for $B^0\to \ell^+\ell^{\prime-}$ at the Tevatron and LHC}},  {\em Phys.Rev.}
  {\bf D79} (2009) 055006, [\href{http://arxiv.org/abs/0812.4320}{{\tt
  arXiv:0812.4320}}].

\bibitem{Hermann:2013kca}
T.~Hermann, M.~Misiak, and M.~Steinhauser, {\it {Three-loop QCD corrections to
  $B_s \to \mu^+ \mu^-$}},  {\em JHEP} {\bf 1312} (2013) 097,
  [\href{http://arxiv.org/abs/1311.1347}{{\tt arXiv:1311.1347}}].

\bibitem{Brod:2011ty}
J.~Brod and M.~Gorbahn, {\it {Next-to-Next-to-Leading-Order Charm-Quark
  Contribution to the CP Violation Parameter $\epsilon_K$ and $\Delta M_K$}},
  {\em Phys.Rev.Lett.} {\bf 108} (2012) 121801,
  [\href{http://arxiv.org/abs/1108.2036}{{\tt arXiv:1108.2036}}].

\bibitem{Christ:2012se}
{\bf RBC and UKQCD} Collaboration, N.~Christ, T.~Izubuchi, C.~Sachrajda,
  A.~Soni, and J.~Yu, {\it {Long distance contribution to the KL-KS mass
  difference}},  {\em Phys.Rev.} {\bf D88} (2013), no.~1 014508,
  [\href{http://arxiv.org/abs/1212.5931}{{\tt arXiv:1212.5931}}].

\bibitem{Bai:2014cva}
Z.~Bai, N.~Christ, T.~Izubuchi, C.~Sachrajda, A.~Soni, {\em et~al.}, {\it
  {$K_L-K_S$ mass difference from lattice QCD}},  {\em Phys.Rev.Lett.} {\bf
  113} (2014) 112003, [\href{http://arxiv.org/abs/1406.0916}{{\tt
  arXiv:1406.0916}}].

\bibitem{Butler:2013kdw}
{\bf Quark Flavor Physics Working Group} Collaboration, J.~Butler {\em et~al.},
  {\it {Working Group Report: Quark Flavor Physics}},
  \href{http://arxiv.org/abs/1311.1076}{{\tt arXiv:1311.1076}}.

\bibitem{Abe:2010gxa}
{\bf Belle-II} Collaboration, T.~Abe {\em et~al.}, {\it {Belle II Technical
  Design Report}},  \href{http://arxiv.org/abs/1011.0352}{{\tt
  arXiv:1011.0352}}.

\bibitem{Aushev:2010bq}
T.~Aushev, W.~Bartel, A.~Bondar, J.~Brodzicka, T.~Browder, {\em et~al.}, {\it
  {Physics at Super B Factory}},  \href{http://arxiv.org/abs/1002.5012}{{\tt
  arXiv:1002.5012}}.

\bibitem{NA62TDR}
{\bf NA62} Collaboration, {\it {NA62 Technical Design Report}},  2010.
\newblock \url{https://na62.web.cern.ch/na62/Documents/TD_Full_doc_v10.pdf}.

\bibitem{Fantechi:2014hqa}
{\bf NA62} Collaboration, R.~Fantechi, {\it {The NA62 experiment at CERN:
  status and perspectives}},  \href{http://arxiv.org/abs/1407.8213}{{\tt
  arXiv:1407.8213}}.

\bibitem{E.T.WorcesterfortheORKA:2013cya}
{\bf ORKA} Collaboration, E.~Worcester, {\it {ORKA, The Golden Kaon Experiment:
  Precision measurement of $K^+ \to \pi^+ \nu \bar{\nu}$ and other rare
  processes}},  {\em PoS} {\bf KAON13} (2013) 035,
  [\href{http://arxiv.org/abs/1305.7245}{{\tt arXiv:1305.7245}}].

\bibitem{Christ:2014qwa}
N.~Christ, T.~Izubuchi, C.~T. Sachrajda, A.~Soni, and J.~Yu, {\it {Calculating
  the $K_L-K_S$ mass difference and $\epsilon_K$ to sub-percent accuracy}},
  {\em PoS} {\bf LATTICE2013} (2014) 397,
  [\href{http://arxiv.org/abs/1402.2577}{{\tt arXiv:1402.2577}}].

\bibitem{Mahbubani:2012qq}
R.~Mahbubani, M.~Papucci, G.~Perez, J.~T. Ruderman, and A.~Weiler, {\it {Light
  Nondegenerate Squarks at the LHC}},  {\em Phys.Rev.Lett.} {\bf 110} (2013),
  no.~15 151804, [\href{http://arxiv.org/abs/1212.3328}{{\tt
  arXiv:1212.3328}}].

\bibitem{Blanke:2013uia}
M.~Blanke, G.~F. Giudice, P.~Paradisi, G.~Perez, and J.~Zupan, {\it {Flavoured
  Naturalness}},  {\em JHEP} {\bf 1306} (2013) 022,
  [\href{http://arxiv.org/abs/1302.7232}{{\tt arXiv:1302.7232}}].

\bibitem{Agrawal:2013kha}
P.~Agrawal and C.~Frugiuele, {\it {Mixing stops at the LHC}},  {\em JHEP} {\bf
  1401} (2014) 115, [\href{http://arxiv.org/abs/1304.3068}{{\tt
  arXiv:1304.3068}}].

\bibitem{Backovic:2015rwa}
M.~Backovi?, A.~Mariotti, and M.~Spannowsky, {\it {Signs of Tops from Highly
  Mixed Stops}},  \href{http://arxiv.org/abs/1504.00927}{{\tt
  arXiv:1504.00927}}.

\bibitem{Park:2010wf}
J.-h. Park, {\it {Metastability bounds on flavour-violating trilinear soft
  terms in the MSSM}},  {\em Phys.Rev.} {\bf D83} (2011) 055015,
  [\href{http://arxiv.org/abs/1011.4939}{{\tt arXiv:1011.4939}}].

\bibitem{Arana-Catania:2014ooa}
M.~Arana-Catania, S.~Heinemeyer, and M.~Herrero, {\it {Updated Constraints on
  General Squark Flavor Mixing}},  {\em Phys.Rev.} {\bf D90} (2014), no.~7
  075003, [\href{http://arxiv.org/abs/1405.6960}{{\tt arXiv:1405.6960}}].

\bibitem{Kowalska:2014opa}
K.~Kowalska, {\it {Phenomenology of SUSY with General Flavour Violation}},
  {\em JHEP} {\bf 1409} (2014) 139, [\href{http://arxiv.org/abs/1406.0710}{{\tt
  arXiv:1406.0710}}].

\bibitem{Pierce:1996zz}
D.~M. Pierce, J.~A. Bagger, K.~T. Matchev, and R.-j. Zhang, {\it {Precision
  corrections in the minimal supersymmetric standard model}},  {\em Nucl.Phys.}
  {\bf B491} (1997) 3--67, [\href{http://arxiv.org/abs/hep-ph/9606211}{{\tt
  hep-ph/9606211}}].

\bibitem{Allanach:2008qq}
B.~Allanach, C.~Balazs, G.~Belanger, M.~Bernhardt, F.~Boudjema, {\em et~al.},
  {\it {SUSY Les Houches Accord 2}},  {\em Comput.Phys.Commun.} {\bf 180}
  (2009) 8--25, [\href{http://arxiv.org/abs/0801.0045}{{\tt arXiv:0801.0045}}].

\bibitem{Aoki:2013ldr}
S.~Aoki, Y.~Aoki, C.~Bernard, T.~Blum, G.~Colangelo, {\em et~al.}, {\it {Review
  of lattice results concerning low-energy particle physics}},  {\em
  Eur.Phys.J.} {\bf C74} (2014), no.~9 2890,
  [\href{http://arxiv.org/abs/1310.8555}{{\tt arXiv:1310.8555}}].

\bibitem{Bae:2013mja}
T.~Bae, Y.-C. Jang, H.~Jeong, J.~Kim, J.~Kim, {\em et~al.}, {\it {Beyond the
  Standard Model B-parameters with improved staggered fermions in $N_f=2+1$
  QCD}},  \href{http://arxiv.org/abs/1310.7372}{{\tt arXiv:1310.7372}}.

\bibitem{Carrasco:2013jaa}
{\bf ETM} Collaboration, N.~Carrasco {\em et~al.}, {\it {$K$ and $D$
  oscillations in the Standard Model and its extensions from $N_f=2+1+1$
  Twisted Mass LQCD}},  \href{http://arxiv.org/abs/1310.5461}{{\tt
  arXiv:1310.5461}}.

\bibitem{Carrasco:2013iba}
{\bf ETM} Collaboration, N.~Carrasco {\em et~al.}, {\it {B-physics computations
  from Nf=2 tmQCD}},  \href{http://arxiv.org/abs/1310.1851}{{\tt
  arXiv:1310.1851}}.

\bibitem{Buras:1998raa}
A.~J. Buras, {\it {Weak Hamiltonian, CP violation and rare decays}},
  \href{http://arxiv.org/abs/hep-ph/9806471}{{\tt hep-ph/9806471}}.

\bibitem{Cirigliano:2011ny}
V.~Cirigliano, G.~Ecker, H.~Neufeld, A.~Pich, and J.~Portoles, {\it {Kaon
  Decays in the Standard Model}},  {\em Rev.Mod.Phys.} {\bf 84} (2012) 399,
  [\href{http://arxiv.org/abs/1107.6001}{{\tt arXiv:1107.6001}}].

\bibitem{Buras:2002tp}
A.~J. Buras, A.~Czarnecki, M.~Misiak, and J.~Urban, {\it {Completing the NLO
  QCD calculation of $\bar B\to X_s\gamma$}},  {\em Nucl.Phys.} {\bf B631}
  (2002) 219--238, [\href{http://arxiv.org/abs/hep-ph/0203135}{{\tt
  hep-ph/0203135}}].

\bibitem{Hurth:2003dk}
T.~Hurth, E.~Lunghi, and W.~Porod, {\it {Untagged $\bar B\to X_{s+d} \gamma$ CP
  asymmetry as a probe for new physics}},  {\em Nucl.Phys.} {\bf B704} (2005)
  56--74, [\href{http://arxiv.org/abs/hep-ph/0312260}{{\tt hep-ph/0312260}}].

\bibitem{Staub:2010jh}
F.~Staub, {\it {Automatic Calculation of supersymmetric Renormalization Group
  Equations and Self Energies}},  {\em Comput.Phys.Commun.} {\bf 182} (2011)
  808--833, [\href{http://arxiv.org/abs/1002.0840}{{\tt arXiv:1002.0840}}].

\bibitem{Marciano:1996wy}
W.~Marciano and Z.~Parsa, {\it {Rare kaon decays with 'missing energy'}},  {\em
  Phys.Rev.} {\bf D53} (1996) 1--5.

\bibitem{Bauer:2004ve}
C.~W. Bauer, Z.~Ligeti, M.~Luke, A.~V. Manohar, and M.~Trott, {\it {Global
  analysis of inclusive B decays}},  {\em Phys.Rev.} {\bf D70} (2004) 094017,
  [\href{http://arxiv.org/abs/hep-ph/0408002}{{\tt hep-ph/0408002}}].

\end{thebibliography}\endgroup
\bibliographystyle{JHEP}

\end{document}